\documentclass[12pt,a4paper]{JHEP3} % 10pt is ignored!

\JHEPspecialurl{http://jhep.sissa.it/JOURNAL/JHEP3.tar.gz}

\usepackage{amsmath}
\usepackage{graphicx}
\usepackage{latexsym}
\usepackage{amsfonts}
\usepackage{amssymb}
\usepackage{epsfig}
\newcommand\fverb{\setbox\fverbbox=\hbox\bgroup\verb}
\newcommand\fverbdo{\egroup\medskip\noindent%
            \fbox{\unhbox\fverbbox}\ }
\newcommand\fverbit{\egroup\item[\fbox{\unhbox\fverbbox}]}
\newbox\fverbbox

\newcommand{\drawsquare}[2]{\hbox{%
\rule{#2pt}{#1pt}\hskip-#2pt%  left vertical
\rule{#1pt}{#2pt}\hskip-#1pt%  lower horizontal
\rule[#1pt]{#1pt}{#2pt}}\rule[#1pt]{#2pt}{#2pt}\hskip-#2pt%  upper horizontal
\rule{#2pt}{#1pt}}% right vertical

\newcommand{\fund}{~\raisebox{-.5pt}{\drawsquare{6.5}{0.4}}~}
\newcommand{\antifund}{~\overline{\raisebox{-.5pt}{\drawsquare{6.5}{0.4}}}~}

\newcommand{\antisym}{~\raisebox{-3.5pt}{\drawsquare{6.5}{0.4}}\hskip-6.9pt%
        \raisebox{3pt}{\drawsquare{6.5}{0.4}}~}%  antisymmetric second rank

\newcommand{\antiasymm}{~\overline{\raisebox{-3.5pt}{\drawsquare{6.5}{0.4}}\hskip-6.9pt%
        \raisebox{3pt}{\drawsquare{6.5}{0.4}}}~}%  antisymmetric second rank

\newcommand{\OO}{\mathcal{O}}

\newcommand{\Tr}{{\rm Tr}}

\newcommand{\be}{\begin{equation}}
\newcommand{\ee}{\end{equation}}
\newcommand{\bea}{\begin{eqnarray}}
\newcommand{\eea}{\end{eqnarray}}

\renewcommand{\t}{\tilde}
\newcommand{\mc}{\mathcal}

\title{Phases of N=1 supersymmetric chiral gauge theories}
\author{Nathaniel Craig$^{1,\,2}$, Rouven Essig$^{3,\,1,\,4}$ ,  Anson Hook$^4$, Gonzalo Torroba$^4$ 
\\
\vspace{0.2cm}
~\\
$^1$ School of Natural Sciences, \\
Institute for Advanced Study, Princeton, NJ 08540, USA \\
\vspace{0.2cm}

$^2$ Department of Physics and Astronomy \\
Rutgers University, Piscataway, NJ 08854, USA \\
\vspace{0.2cm}

$^3$ C.N.~Yang Institute for Theoretical Physics, \\
Stony Brook University, Stony Brook, NY 11794 USA\\
\vspace{0.2cm}

$^4$ Physics Department and SLAC National Accelerator Laboratory \\
Stanford University, CA 94309, USA \\
\vspace{0.2cm}

}

\abstract{
We analyze the phases of supersymmetric chiral gauge theories with an antisymmetric tensor and (anti)fundamental flavors, in the presence of a classically marginal superpotential deformation. Varying the number of flavors that appear in the superpotential reveals rich infrared chiral dynamics and novel dualities. The dualities are characterized by an infinite family of magnetic duals with arbitrarily large gauge groups describing the same fixed point, correlated with 
arbitrarily large classical global symmetries that are truncated nonperturbatively.
At the origin of moduli space, these theories exhibit a phase with confinement and chiral symmetry breaking, an interacting nonabelian Coulomb phase, and phases where an interacting sector coexists with a sector that either s-confines or is in a free magnetic phase. Properties of these intriguing ``mixed phases'' are studied in detail using duality and $a$-maximization, and the presence of superpotential interactions provides further insights into their formation.
}

\preprint{SLAC-PUB-14649 \\ RUNHETC-2011-20 \\ YITP-SB-11-35}

\begin{document}

\tableofcontents

%%%%%%%%%%%%%%%%%%%%%%%%%%%%%%%%%%%%%%%%%%%
%%%%%%%%%%%%%%%%%%%%%%%%%%%%%%%%%%%%%%%%%%%
%%%%%%%%%%%%%%%%%%%%%%%%%%%%%%%%%%%%%%%%%%%
%%%%%%%%%%%%%%%%%%%%%%%%%%%%%%%%%%%%%%%%%%%
\section{Introduction and summary of results}\label{sec:intro}

Supersymmetric chiral gauge theories are theoretically and phenomenologically interesting, not least for their resemblance to the Standard Model. Such theories were the first to exhibit dynamical supersymmetry breaking (beginning with~\cite{Affleck:1984uz,Affleck:1984xz}), and have provided an arena where intriguing dualities and nonperturbative effects have been discovered.\footnote{These include chiral-nonchiral dualities~\cite{Pouliot:1995zc,Pouliot:1996zh}, mixed phases~\cite{Pouliot:1995me,Terning:1997jj,Csaki:2004uj}, and new phase transitions between conformal and confining theories. Various other examples have been studied for instance in~\cite{Intriligator:1995ax,Brodie:1996xm}.} However, while there has been striking progress in understanding vector-like $\mc N=1$ theories using Seiberg duality~\cite{Seiberg:1994bz,Intriligator:1995au}, the situation with chiral theories is much more subtle because a systematic duality procedure is lacking.

The aim of this work is to determine the phase structure of $\mc N=1$ chiral gauge theories with an antisymmetric tensor and to find dual descriptions that capture the long distance dynamics in simple ways. Our approach will be based on the deconfinement method of Berkooz~\cite{Berkooz:1995km} and the recent results of~\cite{CEHT}, where the deconfinement approach was put on a firmer footing and novel properties of supersymmetric chiral gauge theories came to light. One new ingredient that will be crucial here is the presence of a classically marginal superpotential, which gives access to new fixed points and dynamical properties of chiral theories. We will see that combining duality with $a$-maximization~\cite{Intriligator:2003jj} will allow us to map the full phase structure of the theory, providing nontrivial consistency checks on the proposed duals.

In more detail, we consider $SU(N)$ super Yang-Mills with $F$ fundamentals, an antisymmetric, and $N+F-4$ antifundamentals (ensuring anomaly cancellation). The marginal interaction is given by coupling an even number, $F_1$,  of antifundamentals $\t Q$ to the antisymmetric $A$: $W= \t Q A \t Q$. We will determine the phase structure and infrared (IR) dynamics of the theory as a function of $(N, F, F_1)$.

Though classically marginal, we find that the superpotential interaction becomes relevant at long distances, affecting the phase of the theory in a dramatic way:
\begin{itemize}
\item When $F_1 \ge 2F -4$, nonperturbative effects cause a runaway and the theory does not exist.
\item For $F_1=2F-6$ the theory confines and breaks chiral symmetry.
\item When $F_1=2F-8$ the theory flows to a superconformal field theory (SCFT) plus an s-confinining subsector.
\item For $F_1 < 2F-8$ the theory is in a nonabelian Coulomb phase at the origin of moduli space.
\end{itemize}
These different regimes will be analyzed in terms of the electric and magnetic theories and via $a$-maximization (when there is an interacting superconformal fixed point).

Our analysis will reveal that decreasing $F_1$ below $2F-8$ also changes the magnetic description in a crucial way: while for $F_1 \ge 2F-8$ the dual has at most a simple gauge group, when $F_1<2F-8$ we find product gauge group theories at the origin of moduli space. Dualities relating theories with simple and non-simple gauge groups at the origin of field space are quite interesting; some applications will be studied in~\cite{future}. We should stress that this product gauge group structure occurs even at the origin of moduli space, so it is not related to higgsing the electric theory. These dual descriptions exhibit three striking features:
\begin{enumerate}
\item A ``mixed phase'' consisting of weakly-interacting matter charged under an IR-free gauge sector, coupled to an interacting conformal sector. 
\item Classical global symmetries in the conformal sector that are truncated by nonperturbative effects.
\item An infinite family of magnetic theories, with arbitrarily large gauge groups, all flowing to the same fixed point.
\end{enumerate}
Depending on the parity of $N$ and $F$ we will uncover a rich phase structure in the infrared. Our analysis will reveal an intricate interplay between gauge dynamics, renormalization-group (RG) evolution, and nonperturbative effects.

Following this change in the IR phase as $F_1$ is varied, the analysis in this paper is divided in two parts: \S \ref{sec:nonCFT} is devoted to the chiral theory in the range $F_1 \ge 2F-8$, while \S \ref{sec:mixed} discusses the dynamics when $F_1< 2F-8$. In \S\ref{sec:amax}, we obtain exact results on the SCFT regime using $a$-maximization and provide further evidence for our proposed dualities.  We finish with some concluding statements in \S\ref{sec:discussion}.  We reserve several useful results for Appendix \ref{sec:decon}, which contains the general duality flow using the deconfinement technique. 

Before proceeding into the specifics of the IR phases, let us first present in more detail the electric theory and provide an overview of our main results.

%%%%%%%%%%%%%%%%%%%%%%%%%%%%%%%%%%%%%%%%%%%
%%%%%%%%%%%%%%%%%%%%%%%%%%%%%%%%%%%%%%%%%%%
\subsection{Electric theory and phase structure}

The theory considered in this paper is
\begin{center}
\be\label{table:uv}
\begin{tabular}{c|c|ccc}
&$SU(N)$&$SU(F_2)$&$Sp(F_1)$&$SU(F)$  \\
\hline
&&&&\\[-12pt]
$Q$&$\fund$&$1$&$1$&$\antifund$  \\
$\t Q$&$\antifund$&$1$&$ \fund$ &$1$ \\
$\t P$&$\antifund$&$\fund$&$1$&$1$  \\
$A$&$\antisym$&$1$&$1$&$1$  \\
\end{tabular}
\ee
\end{center}
with superpotential
\be\label{eq:Wchiral1}
W_\text{el}=  \t Q A \t Q\,.
\ee
Anomaly cancellation requires 
\be
F_2 \equiv N+F-F_1-4\,.
\ee
Our goal is to understand the phase structure and IR dynamics of this theory as a function of $(N, F, F_1)$.\footnote{ The theory with $F = N+3, F_1 = 2N -2, F_2 = 1$ was studied in \cite{CEHT}. In general, we will be interested in arbitrary values of $F$ and $F_1$. The theory with $F_1 = 0$ corresponds to $W_\text{el} = 0$, i.e., the chiral theory with no superpotential.}  In particular, varying the number $F_1$ of flavors with superpotential interactions gives a new handle on the formation of mixed phases and will provide further insights into these mysterious structures.

The chiral ring is parameterized by the mesons
\be\label{eq:mesons}
M= \t P Q\;,\;R= \t Q Q\;,\;H= \t P A \t P
\ee
and baryons
\be\label{eq:baryons1}
\t P^{N}\;,\;QA^{(N-1)/2}\;,\;Q^3 A^{(N-3)/2}\;,\;\ldots ,\,Q^k A^{(N-k)/2}\;\;(\text{for}\;N\;\text{odd})
\ee
\be\label{eq:baryons2}
\t P^{N}\;,\;A^{N/2}\;,\;Q^2 A^{(N-2)/2}\;,\;\ldots ,\,Q^k A^{(N-k)/2}\;\;(\text{for}\;N\;\text{even})\,.
\ee
Here $k \le {\rm min}(N,F)$.

\begin{figure}[htb]
\centering
\includegraphics[width=4in]{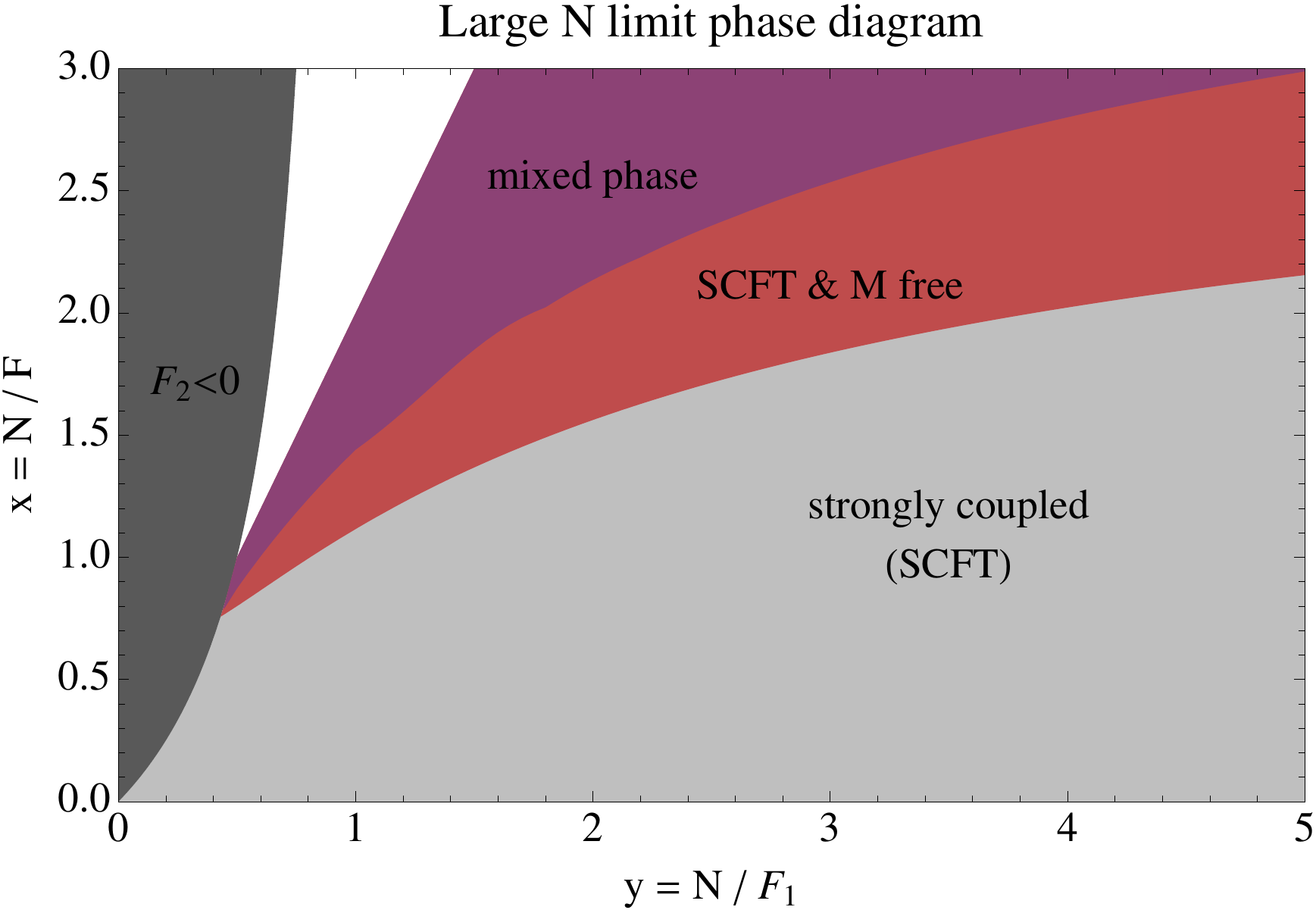}
\caption{\small{Phase diagram of the chiral theory at large $N$, where $x=N/F$ and $y=N/F_1$. In the light grey region we have a superconformal fixed point and all the fields are strongly coupled. $M$ becomes free in the red region, although the theory is still strongly coupled. In the boundary with the purple region, $H$ also becomes free. Beyond this point there are additional $U(1)$ symmetries that cannot be captured in the electric theory. The purple region corresponds to a mixed phase.  In the dark grey region, $F_2<0$, which is physically inaccessible.  The theory is not superconformal in the white region and, approaching the purple region, either has no stable vacuum, confines with chiral symmetry breaking, or s-confines.}}
\label{fig:phase}
\end{figure}

Although we will eventually be interested in dual descriptions for this theory, we may also explore the phase diagram of the electric theory directly. The most striking phase structure occurs in the range $F_1< 2F-8$, in which the theory flows to a superconformal fixed point that may be studied using $a$-maximization (see \S \ref{sec:amax}).\footnote{The nonperturbative effects and phase structure for $F_1 \ge 2F-8$ will be discussed in detail in \S \ref{sec:nonCFT}.} The resulting phase diagram in the large N limit is sketched in Fig.~\ref{fig:phase}. At large $N$, the phase structure is determined by the parameters
\be
x \equiv \frac{N}{F}\;,\;y \equiv \frac{N}{F_1}\,.
\ee
For small $x$ and large $y$ (light grey region), $a$-maximization reveals that all the fields are strongly coupled at the IR fixed point. As we increase $x$ and/or decrease $y$, the first gauge invariant to become free is the meson $M =\t P Q$. The red region defines the open range for $(x,y)$ where $M$ is the only free field (see \S \ref{sec:amax} for more details).

Eventually, the meson $H = \t P A \t P$ also hits the unitarity bound and becomes free. This defines the boundary with the purple region. Beyond this point we expect additional gauge invariants to become free. 
%; this is suggested for instance by the particular case $N=3$, where $H$ is a baryon.  
However, it is not clear how to correct the $a$-maximization calculation to include these effects -- a magnetic dual description of the fixed point is needed. It turns out that using this magnetic description, the purple region is characterized by the appearance of a mixed phase, as we summarize below.

The two-dimensional phase diagram of Fig.~\ref{fig:phase} is made possible by the nonzero superpotential, and sheds light on important aspects of supersymmetric chiral dynamics. Importantly for our purposes, it allows us to approach the mixed phase (purple region) from different limits. Inside the white region and increasing $y$ towards the mixed phase, we will first encounter a theory with a runaway instability, then confinement with chiral symmetry breaking, and for $F_1=2F-8$ (the straight line boundary between the white and purple region) an interacting fixed point plus an s-confining sector. The magnetic description reveals that a further increase in $y$ turns this s-confining sector into a full free magnetic sector, decoupled from the nontrivial fixed point. A similar phase transition is observed in vector-like SQCD, so this provides a physical explanation for the formation of mixed phases, at least in a given duality frame.

On the other hand, we will analyze the chiral dynamics starting from the light grey region (where the theory is at an interacting fixed point) and increasing $x$ towards the mixed phase. In this case, first $M$ becomes free, and then $H$ and an entire free magnetic sector decouple. Also, while $a$-maximization cannot be applied to the electric theory inside the mixed phase region, we do find that $a$-maximization on the electric and magnetic descriptions gives the same results on both boundaries of this region.

%%%%%%%%%%%%%%%%%%%%%%%%%%%%%%%%%%%%%%%%%%%
%%%%%%%%%%%%%%%%%%%%%%%%%%%%%%%%%%%%%%%%%%%
\subsection{Magnetic duals}

Of course, we may gain additional insight by obtaining a magnetic description that is dual to the electric theory. We will find a variety of novel dual descriptions that flow to the same IR fixed point, each of which is characterized by a product gauge group theory. The first step is to obtain a magnetic dual valid for arbitrary $F_1$; the simplest dual description is given in (\ref{table:interm1}), and more general duals appear in (\ref{table:interm}). This duality will be used in \S \ref{sec:nonCFT} to understand the nonperturbative effects and IR dynamics for $F_1 \ge 2F-8$.

On the other hand, for $F_1 < 2F-8$ the magnetic theory can be further dualized, as explained in \S \ref{sec:mixed}. 
These theories will have the feature (inherited from Seiberg duality) that the gauge invariants becoming free are elementary fields; this simplifies the $a$-maximization calculation considerably. In a region of parameter space $(N, F, F_1)$ (the purple region of Fig.~\ref{fig:phase}), the fixed point theory consists of a sector of weakly-interacting fields coupled to a conformal gauge theory.  The weakly-interacting sector is charged under an IR-free $Sp(2F-8-F_1)$ gauge symmetry and interacts with the conformal theory via both bifundamental matter and irrelevant interactions. 

The details of the conformal sector depend on the parities of $N$ and $F$. When $F$ is even, the conformal sector is manifestly self-dual for either parity of $N$. In particular, for {\it odd} $N$ this self-dual theory possesses the same global symmetries as the chiral theory. For {\it even} $N$ the self-dual theory possesses an additional classical $SU(2)$ global symmetry that is truncated by nonperturbative effects. 

\begin{figure}[!t] %  figure placement: here, top, bottom, or page
   \centering
   \includegraphics[width=4in]{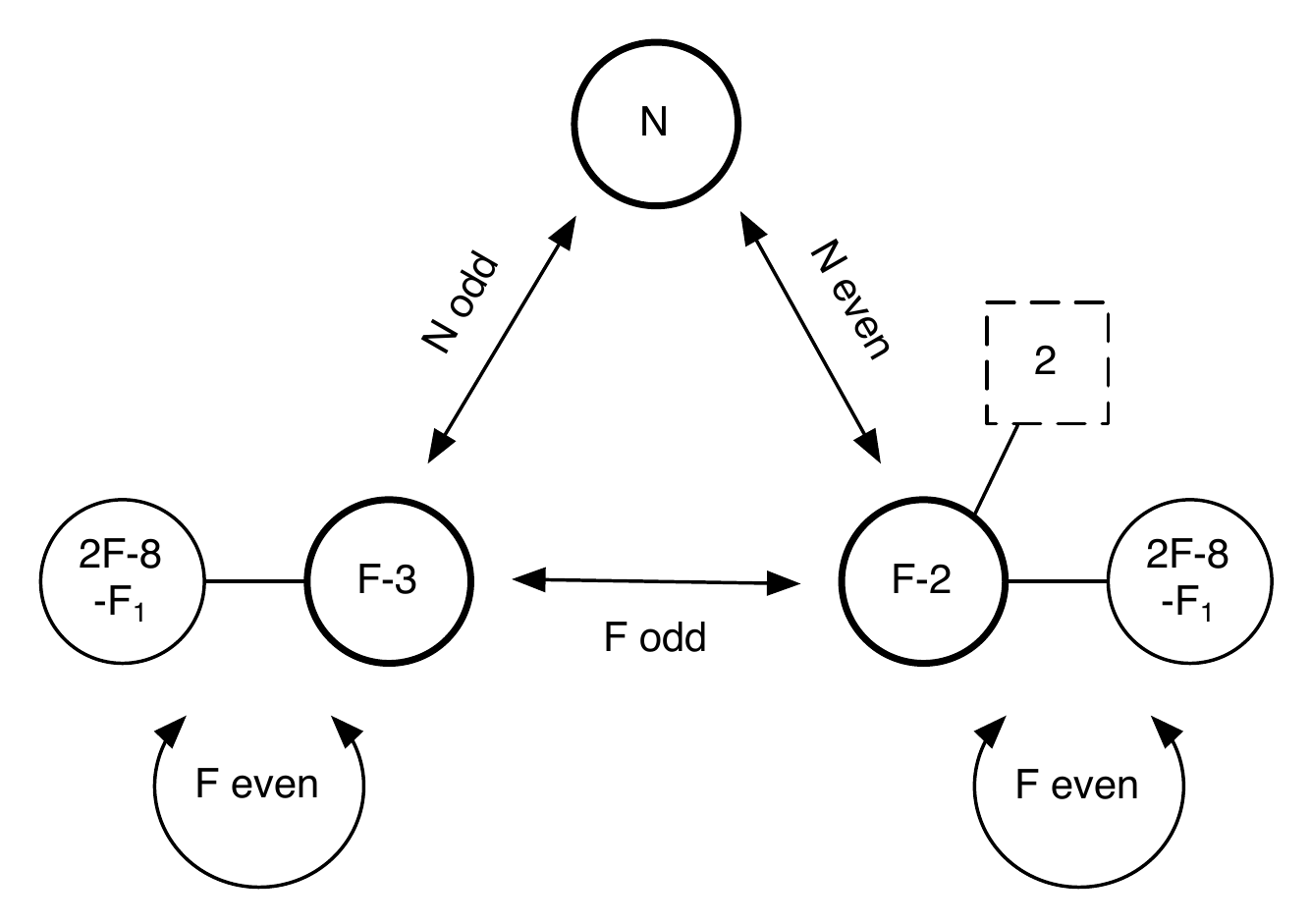} 
   \caption{A sketch of the simplest duality map. The chiral theory for odd $N$ is dual to a $SU(F-3) \times Sp(2F-8-F_1)$ theory. For $F$ even this magnetic theory is self-dual, while for $F$ odd the magnetic theory is dual to a $SU(F-2) \times Sp(2F-8-F_1)$ theory with an additional $SU(2)$ global symmetry that is truncated by nonperturbative effects. Similarly, the chiral theory for even $N$ is dual to a $SU(F-2) \times Sp(2F-8-F_1)$ theory with a truncated $SU(2)$. For $F$ even this theory is self-dual, while for $F$ odd it is dual to an  $SU(F-3) \times Sp(2F-8-F_1)$ theory. All theories possess an additional $SU(F) \times  SU(N+F-F_1-4) \times Sp(F_1)$ global symmetry that has been omitted for clarity.}
   \label{fig:map}
\end{figure}

In contrast, when $F$ is odd, the conformal sector possesses two dual descriptions for either parity of $N$. One dual has the same global symmetries as the chiral theory, while the other involves an additional truncated $SU(2)$ global symmetry. This intricate map of dualities is sketched in Fig.~\ref{fig:map}. The crucial feature of all these dual descriptions, whatever the parity of $N$ and $F$, is that they possess the same quantum global symmetries as (\ref{table:uv}), providing nontrivial tests on the dualities.

In fact, these dual descriptions are the simplest exemplars of an {\it infinite} family of dual theories parameterized by additional classical global symmetries and arbitrarily large magnetic gauge groups. We will find that all the additional classical degrees of freedom associated with these symmetries are truncated quantum mechanically.

%%%%%%%%%%%%%%%%%%%%%%%%%%%%%%%%%%%%%%%%%%%
%%%%%%%%%%%%%%%%%%%%%%%%%%%%%%%%%%%%%%%%%%%
%%%%%%%%%%%%%%%%%%%%%%%%%%%%%%%%%%%%%%%%%%%
%%%%%%%%%%%%%%%%%%%%%%%%%%%%%%%%%%%%%%%%%%%
\section{Chiral theory in the range $F_1 \ge 2F-8$}\label{sec:nonCFT}

This section presents the phase structure of the theory in the range $F_1 \ge 2F-8$. First we analyze nonperturbative effects directly in the electric theory. Using holomorphy and symmetries, the nonperturbative superpotential is obtained (up to an overall coefficient), thus predicting new instanton effects in chiral theories. Next we obtain a magnetic dual that is valid for all $F_1$. For $F_1 \ge 2F-8$ the magnetic description offers an explicit way of deriving the nonperturbative effects and IR phase structure. When $F_1 < 2F-8$ further dualities are possible, leading to the superconformal theories of \S \ref{sec:mixed}.

%%%%%%%%%%%%%%%%%%%%%%%%%%%%%%%%%%%%%%%%%%%
%%%%%%%%%%%%%%%%%%%%%%%%%%%%%%%%%%%%%%%%%%%
\subsection{Nonperturbative effects in the electric theory}\label{eq:npel}

The basic properties of the phase structure for $F_1 \ge 2F-8$ can already be seen in the electric theory, using holomorphy and symmetries. For this purpose, we will consider the following anomalous symmetries:
\begin{center}
\be \label{tab:anom1}
\begin{tabular}{c|ccc}
&$U(1)_Q$&$U(1)_{\t P}$&$U(1)_A$   \\
\hline
&&&\\[-12pt]
$Q$&$1$&$0$&$0$  \\
$\t Q$&$0$&$0$&$-1$  \\
$\t P$&$0$&$1$&$0$  \\
$A$&$0$&$0$&$2$  \\
$\Lambda^{2N-F+3}$&$F$&$N+F-F_1-4$&$2N-4-F_1$  \\
\end{tabular}
\ee
\end{center}
where $\Lambda$ is the dynamical scale of the electric theory.

The dynamical superpotential must be constructed out of chiral ring operators.  The only chiral operator which involves $\t Q$ is the meson $Q \t Q$.  However, this meson can not be made gauge and flavor invariant so the field $\t Q$ cannot appear in the dynamical superpotential.  Thus the generated superpotential is of the schematic form
\be
W_\text{dyn}=Q^a \t P^b A^c (\Lambda^{2N-F+3})^d  \,.
\ee
This superpotential must be invariant under the three $U(1)$ symmetries and have classical dimension 3.  For $F_1> 2F-6$, the solution to these equations gives
\be\label{eq:Wnp1}
W_\text{dyn}=C_{N,F,F_1}\left(\frac{\Lambda^{2N-F+3}}{Q^F \t P^{N+F-F_1-4} A^{\frac{2 N- F_1-4}{2}}}\right)^\frac{2}{F_1-(2F-6)}\,,
\ee
with $C_{N,F,F_1}$ is an arbitrary constant. When $F_1=2F-6$, holomorphy and symmetries are consistent with the existence of a constraint,
\be\label{eq:constr-el}
C_{N,F,2F-6}\left(Q^F \t P^{N-F+2} A^{N-F+1}-\Lambda^{2N-F+3} \right)=0\,,
\ee 
corresponding to a quantum modified moduli space.

An important conclusion from this analysis is that for $F_1 \le 2F-8$ it is not possible to generate a dynamical superpotential. Indeed, in this range (\ref{eq:Wnp1}) does not have a sensible weak-coupling limit $\Lambda \to 0$. This is reminiscent of the vanishing of the dynamical superpotential in the magnetic description of SQCD in the conformal window. So already at this stage we see indications that decreasing $F_1$ to $2F-8$ or below may change the phase structure in important ways. It is also necessary to point out that in the present approach, the constants $C_{N,F,F_1}$ are not fixed. Using the magnetic duals below, it will be shown that $C_{N,F,F_1} \neq 0$. It would be interesting to check these predictions for dynamical effects in terms of instanton calculations in the electric theory.

Note that the superpotential (\ref{eq:Wnp1}) may be written in terms of different combinations of gauge invariants. Indeed, recalling (\ref{eq:mesons})--(\ref{eq:baryons2}), we obtain the combination of mesons and baryons
\be\label{eq:combi}
Q^F \t P^{N+F-F_1-4} A^{\frac{2 N- F_1-4}{2}} \sim M^n \left(Q^{F-n} A^{(N-F-n)/2} \right) H^{(F_2-n)/2}
\ee
where $F_2=N+F-F_1-4$, and $n$ is any positive integer with the same parity as $N-F$. We will see shortly that the same structure arises in the magnetic dual.

%%%%%%%%%%%%%%%%%%%%%%%%%%%%%%%%%%%%%%%%%%%
%%%%%%%%%%%%%%%%%%%%%%%%%%%%%%%%%%%%%%%%%%%
\subsection{A magnetic description for arbitrary $F_1$}

The basic tool employed in this work to derive new dualities is the deconfinement method of \cite{Berkooz:1995km}, later  generalized by \cite{Luty:1996cg}. This is described in detail in Appendix \ref{sec:decon}. The dual that we present here follows from the intermediate step (\ref{table:interm}). Here we focus on the case of odd $N$ and set $K=1$; the cases of even $N$ and arbitrary $K$ are further discussed in \S \ref{sec:mixed}.

Based on these results, the proposed magnetic dual is
\begin{center}
\be\label{table:interm1}
\begin{tabular}{c|cc|ccc}
&$SU(F-3)$&$Sp(N-3)$&$SU(F_2)$&$Sp(F_1)$&$SU(F)$   \\
\hline
&&&&\\[-12pt]
$q$&$\fund$&$1$&$1$&$1$&$\fund$  \\
$\t q$&$\antifund$&$1$&$1$&$ \fund$ &$1$ \\
$\t p$&$\antifund$&$1$&$\antifund$&$1$&$1$  \\
$x$&$ \fund$&$\fund$&$1$&$1$ &$1$ \\
$y$&$1$&$\fund$&$\fund$&$1$ &$1$ \\
$u$&$\antifund$&$1$&$1$&$ 1$ &$1$ \\
$R$&$1$&$1$&$1$&$\fund$ &$\antifund$ \\
$M$&$1$&$1$&$\fund$&$1$ &$\antifund$ \\
$S$&$1$&$1$&$1$&$1$ &$\antifund$
\end{tabular}
\ee
\end{center}
with a superpotential
\be\label{eq:Winterm1}
W= \t q xx \t q+q R \t q+q M \t p+q S u+x y \t p\,.
\ee
The matching of operators can be obtained using (global) abelian and nonabelian symmetries. See \cite{CEHT} for more details in a related context. For later purposes, we simply note that
\bea\label{eq:matching}
H & \leftrightarrow & (y y)\nonumber\\
Q^{F-n} A^{(N-F-n)/2} & \leftrightarrow & q^n (xx)^{(F-3-n)/2}\,.
\eea

In this duality frame, the nonperturbative effects are caused predominantly by the $Sp$ dynamics. The $Sp$ gauge group has $2N_c \equiv N-3$ colors and $2 N_f \equiv (2F-4-F_1)+ 2N_c$ flavors. The basic gauge and flavor invariant built from the $Sp$ mesons $\mc M = \{(xx), (xy), (y y) \}$ is
\be
\text{Pf}\,\mc M = \sum_n\,c_n\,(xx)^{(F-3-n)/2} (x y)^n (y y)^{(F_2-n)/2}
\ee
where $c_n$ are combinatorial factors from the pfaffian.

We now determine the phase structure after the $Sp$ factor becomes strong, using the results of \cite{Intriligator:1995ne}.

%%%%%%%%%%%%%%%%%%%%%%%%%%%%%%%%%%%%%%%%%%%
%%%%%%%%%%%%%%%%%%%%%%%%%%%%%%%%%%%%%%%%%%%
\subsection{Runaway instability for $F_1 > 2F-6$}

Considering first  $F_1 > 2F-6$, the $Sp$ dynamics has $N_f<N_c+1$. This leads to a runaway superpotential
\be
W_\text{dyn} = \left(\frac{\Lambda_{Sp}^{(3N-F-F_2)/2}}{\sum_n\,c_n\,(xx)^{(F-3-n)/2} (x y)^n (y y)^{(F_2-n)/2}}\right)^{\frac{2}{F_1-(2F-6)}}\,.
\ee
Combining with the superpotential (\ref{eq:Winterm1}) and setting $W_{\t p}=0$ gives
\be\label{eq:Wnp2}
W= \t q (xx) \t q+ q R \t q + q S u + \left(\frac{\Lambda_{Sp}^{(3N-F-F_2)/2}}{\sum_n\,c_n\,M^n \left( q^n(xx)^{(F-3-n)/2}\right) (y y)^{(F_2-n)/2}}\right)^{\frac{2}{F_1-(2F-6)}}\,.
\ee
This theory displays a runaway behavior without a stable vacuum.

Recalling the operator matching (\ref{eq:matching}) together with (\ref{eq:combi}), we see that the two dynamical superpotentials in (\ref{eq:Wnp1}) and (\ref{eq:Wnp2}) agree, and $C_{N,F,F_1} \neq 0$. Thus we have reproduced the nonperturbative effects of the electric theory in terms of $Sp$ instanton effects via deconfinement.

%%%%%%%%%%%%%%%%%%%%%%%%%%%%%%%%%%%%%%%%%%%
%%%%%%%%%%%%%%%%%%%%%%%%%%%%%%%%%%%%%%%%%%%
\subsection{$F_1=2F-6$: confinement and chiral symmetry breaking}

Next, when $F_1=2F-6$ the $Sp$ factor confines and breaks chiral symmetry. The constraint in the magnetic theory reads
\be
\sum_n\,c_n\,M^n \left( q^n a^{(F-3-n)/2}\right) H^{(F_2-n)/2}=\Lambda_{Sp}^{F+F_2-3}\,,
\ee
where we have written $a \equiv (xx)$ (an antisymmetric under $SU(F-3)$) and $H \equiv (y y)$ (an antisymmetric of $SU(F_2=N-F+2)$). This is consistent with the electric description result (\ref{eq:constr-el}). Let us discuss the simplest $n=0$ branch, where $a^{(F-3)/2}H^{F_2/2}\neq 0$.

After confinement, $(x y)$ and $\t p$ become massive. The nonzero expectation value of $a$ breaks the gauge group $SU(F-3) \to Sp(F-3)$, while $H$ breaks the global subgroup $SU(F_2) \to Sp(F_2)$. The VEV  of $a$ also induces a mass term for $\t q$. The low energy theory then becomes
\begin{center}
\be
\begin{tabular}{c|c|ccc}
&$Sp(F-3)$&$Sp(N-F+2)$&$Sp(2F-6)$&$SU(F)$   \\
\hline
&&&&\\[-12pt]
$q$&$\fund$&$1$&$1$&$\fund$  \\
$u$&$\antifund$&$1$&$ 1$ &$1$ \\
$R$&$1$&$1$&$\fund$ &$\antifund$ \\
$M$&$1$&$\fund$&$1$ &$\antifund$ \\
$H$&$ 1$&$\antisym+1$&$1$ &$1$ \\
$S$&$1$&$1$&$1$ &$\antifund$
\end{tabular}
\ee
\end{center}
with, schematically, 
\be
W= qR qR + qu S\,.
\ee
(Here we have been a bit simplistic with the low energy spectrum, which in fact could come from different combinations of the original fields before symmetry breaking. It may also be interesting to study in more detail the other branches with $n >0$.)

Now the remaining $Sp(F-3)$ group s-confines, producing mesons $(qq)$ and $(qu)$. The latter becomes massive through the coupling to $S$. We conclude that the dual of the electric theory with $F_1=2F-6$ is a weakly coupled theory of gauge singlets, with global symmetries
\begin{center}
\be \label{tab:dual2F6}
\begin{tabular}{c|ccc}
&$Sp(N-F+2)$&$Sp(2F-6)$&$SU(F)$   \\
\hline
&&&\\[-12pt]
$(qq)$&$1$&$1$&$\antisym$  \\
$R$&$1$&$\fund$ &$\antifund$ \\
$M$&$\fund$&$1$ &$\antifund$ \\
$H$&$\antisym+1$&$1$ &$1$ \\
\end{tabular}
\ee
\end{center}
and interactions
\be
W= R (qq) R\,.
\ee

%%%%%%%%%%%%%%%%%%%%%%%%%%%%%%%%%%%%%%%%%%%
%%%%%%%%%%%%%%%%%%%%%%%%%%%%%%%%%%%%%%%%%%%
\subsection{$F_1 = 2F-8$: CFT plus s-confinement}

Finally, when $F_1=2F-8$ the $Sp(N-3)$ factor in (\ref{table:interm1}) s-confines. Integrating out the massive fields (which also sets the nonperturbative superpotential to zero), we arrive at the magnetic description
\begin{center}
\be \label{tab:ir1-sconf}
\begin{tabular}{c|c|ccc}
&$SU(F-3)$&$SU(N-F+4)$&$Sp(2F-8)$&$SU(F)$   \\
\hline
&&&&\\[-12pt]
$q$&$\fund$&$1$&$1$&$\fund$  \\
$\t q$&$\antifund$&$1$&$ \fund$ &$1$ \\
$u$&$\antifund$&$1$&$ 1$ &$1$ \\
$a$&$ \antisym$&$1$&$1$ &$1$ \\
$R$&$1$&$1$&$\fund$ &$\antifund$ \\
$S$&$1$&$1$&$1$ &$\antifund$ \\
\hline
&&&&\\[-12pt]
$H$&$ 1$&$\antisym$&$1$ &$1$ \\
$M$&$1$&$\fund$&$1$ &$\antifund$ \\
\end{tabular}
\ee
\end{center}
with superpotential
\be\label{Wsconf}
W=\t q a \t q+q R \t q+q S u\,.
\ee
So we conjecture that the $SU(N)$ electric theory with $F_1=2F-8$ has a dual magnetic description in terms of a $SU(F-3)$ theory with matter content (\ref{tab:ir1-sconf}).

The magnetic theory consists of a strongly coupled SCFT plus the free gauge singlets $M$ and $H$. It is interesting to note that the strongly coupled subsector is independent of $N$ -- a property that is hard to anticipate in the $SU(N)$ electric theory. The IR dynamics can also be studied directly in the electric theory using $a$-maximization, providing an independent check on the duality. This reveals, in agreement with the magnetic dual predictions, that the mesons $H$ and $M$ become free, while the rest of the fields have nonzero anomalous dimensions independent of $N$. Also, for this value of $F_1$ the nonperturbative superpotential from the $Sp(N-3)$ factor is set to zero by F-term constraints. This provides a direct proof of the vanishing of the dynamical superpotential, in agreement with \S \ref{eq:npel}.

This ends our analysis of the chiral theory in the range $F_1 \ge 2F-8$; in the next section we will study the range  $F_1 < 2F-8$. One of our main results will be that the field $H$ (produced by the s-confinement of $Sp(N-3)$) is upgraded to a whole new sector, with its own gauge dynamics and quarks. This sector may be weakly or strongly coupled (depending on the values of $(N,F,F_1)$), and interacts with the $SU(F-3)$ part that we already found. The enhancement of $H$ to a whole new gauge sector is analogous to what happens in vector-like SQCD as we change between $N_f=N_c+1$ and $N_f>N_c+1$.

%%%%%%%%%%%%%%%%%%%%%%%%%%%%%%%%%%%%%%%%%%%
%%%%%%%%%%%%%%%%%%%%%%%%%%%%%%%%%%%%%%%%%%%
%%%%%%%%%%%%%%%%%%%%%%%%%%%%%%%%%%%%%%%%%%%
%%%%%%%%%%%%%%%%%%%%%%%%%%%%%%%%%%%%%%%%%%%
\section{Chiral theory in the range $F_1 < 2 F-8$  }\label{sec:mixed}

Let us now study the chiral theory for $F_1<2F-8$. At the origin of moduli space, this theory is in an interacting 
non-Abelian Coulomb phase. Our interest will be in its low-energy behavior, which can be understood in a 
dual theory that is more tractable. In this section, we will present the dual theory, 
and show that at low energies (and for large enough $N$ and $F_1$) the theory  exists in a ``mixed phase'' --- a phase consisting of two gauge groups, one of which is at an interacting IR conformal fixed point and the other of which is IR free.

Mixed phases in chiral theories were first observed in~\cite{Terning:1997jj,Csaki:2004uj}, where the matter content considered here was analyzed with $N$ odd and a vanishing superpotential. The case $N$ even has been considered with the non-zero superpotential 
$W = {\rm Pf}(A)$ in~\cite{Berkooz:1995km} and with a vanishing superpotential 
in~\cite{Pouliot:1995me,Csaki:2004uj}. However, as we will show, significant further progress may be made in understanding the IR dynamics. The limit $F_1 = 0$ reproduces the results of~\cite{Pouliot:1995me,Terning:1997jj,Csaki:2004uj}, albeit with an improved understanding of the global symmetries.

%%%%%%%%%%%%%%%%%%%%%%%%%%%%%%%%%%%%%%%%%%%
%%%%%%%%%%%%%%%%%%%%%%%%%%%%%%%%%%%%%%%%%%%
\subsection{The theory for $N$ odd}\label{subsec:odd}

Let us begin with the chiral theory of (\ref{table:uv}) for $N$ odd and $F_1<2F-8$. The theory with $N$ even exhibits additional novel features and will be treated in \S\ref{subsec:even}.

%%%%%%%%%%%%%%%%%%%%%%%%%%%%%%%%%%%%%%%%%%%
\subsubsection{Magnetic description}

When $N$ is odd, a dual description for (\ref{table:uv}) may be obtained via deconfinement.\footnote{For conciseness, we will typically omit the explicit duality steps of deconfinement; the deconfinement procedure for the most general theory is presented in the Appendix.} The dual theory consists of an $SU(F-3) \times Sp(2F-8-F_1)$ gauge theory with matter 
\begin{center}
\be \label{tab:ir1}
\begin{tabular}{c|cc|cccc}
&$SU(F-3)$&$Sp(2F-8-F_1)$&$SU(F_2)$&$Sp(F_1)$&$SU(F)$   \\
\hline
&&&&&\\[-12pt]
$q$&$\fund$&$1$&$1$&$1$&$\fund$  \\
$\t q$&$\antifund$&$1$&$1$&$ \fund$ &$1$ \\
$u$&$\antifund$&$1$&$1$&$ 1$ &$1$ \\
$\t x$&$\antifund$&$\fund$&$1$&$ 1$ &$1$ \\
$a$&$ \antisym$&$ 1$&$1$&$1$ &$1$ \\
$R$&$1$&$1$&$1$&$\fund$ &$\antifund$ \\
$S$&$1$&$1$&$1$&$1$ &$\antifund$ \\
\hline
&&&&\\[-12pt]
$l$&$1$&$\fund$&$\antifund$&$ 1$ &$1$ \\
$H$&$ 1$&$ 1$&$\antisym$&$1$ &$1$ \\
$M$&$1$&$1$&$\fund$&$1$ &$\antifund$ \\
\end{tabular}
\ee
\end{center}
and the superpotential is
\be
W=\t q a \t q+q R \t q+q M \t x l+q S u
+\t x a \t x +l H l\,.
\ee
We remind the reader that $F_1+ F_2= N+F-4$.

Intriguingly, for large enough $N$ and $F_1$ this dual theory factorizes into two distinct sectors, in a way that will be made precise in \S\ref{sec:amax}:
\begin{itemize}
\item An IR free $Sp(2F-8-F_1)$ gauge group with fundamental matter $\t x, l$ and gauge-singlet mesons $H, M$. 
\item An interacting $SU(F-3)$ gauge group with fundamental matter $q, \t q, u$, $\t x$, an antisymmetric tensor $a$, and gauge-singlet mesons $R,S$.
\end{itemize}
 Notice that the rank of the IR free subsector depends on the number $F_1$ of flavors that interact cubicly with the antisymmetric. Increasing the number of interaction terms has the effect of decreasing the size of the free subsector.

\begin{figure}[t] %  figure placement: here, top, bottom, or page
   \centering
   \includegraphics[width=4in]{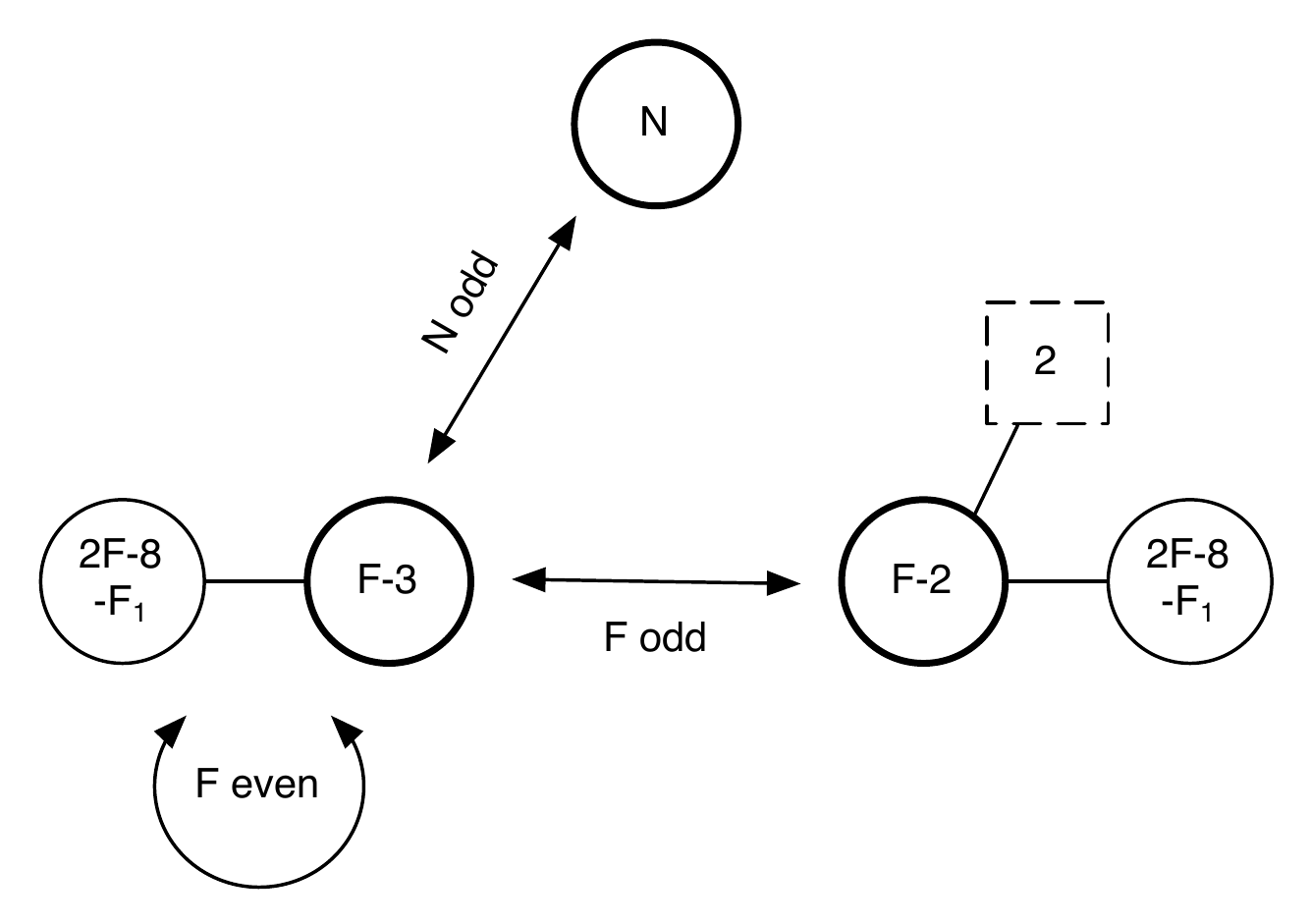} 
   \caption{A sketch of the duality map for odd $N$. The chiral theory is dual to a $SU(F-3) \times Sp(2F-8-F_1)$ theory. For $F$ even this magnetic theory is self-dual, while for $F$ odd the magnetic theory is dual to a $SU(F-2) \times Sp(2F-8 - F_1)$ theory with an additional $SU(2)$ global symmetry that is truncated by nonperturbative effects. All theories possess an additional $SU(F) \times SU(N+F-F_1-4) \times Sp(F_1)$ global symmetry that has been omitted for clarity.}
   \label{fig:odd}
\end{figure}

The two sectors are connected by the bifundamental field $\t x$ and the perturbatively irrelevant operator $q l M \t x$. One might naturally worry that the large anomalous dimensions of fields in the interacting sector would spoil the IR freedom of the $Sp(2F-8-F_1)$ sector. However, for a given $F$ and $F_1$ we may always find a value of $N$ such that the $Sp(2F-8-F_1)$ group runs free and the operator $q l M \t q$ is truly irrelevant. In this case we may treat the two sectors as factorized at low energies. We will make this more concrete using $a$-maximization in \S \ref{sec:amax}.

For sufficiently many flavors the $Sp(2F-8-F_1)$ sector goes free in the infrared, and its dynamics are therefore well-understood. More interesting are the infrared dynamics of the $SU(F-3)$ sector. As we will show, the interacting $SU(F-3)$ sector is at a {\it self-dual} conformal fixed point in the infrared for even $F$. For odd $F$, the dual consists of 
an $SU(F-2)$ magnetic gauge group with additional truncated global symmetries (see Fig.~\ref{fig:odd}).

To see this, it is useful to treat the free $Sp(2F-8-F_1)$ sector as a global symmetry of the $SU(F-3)$ sector. For clarity, we will set aside the free fields associated with the $Sp(2F-8-F_1)$ sector. The interacting degrees of freedom in the magnetic theory are then simply

\begin{center}
\be  \label{tab:simpleir1}
\begin{tabular}{c|c|cccc}
&$SU(F-3)$&$Sp(2F-8-F_1)$&$Sp(F_1)$&$SU(F)$   \\
\hline
&&&&&\\[-12pt]
$q$&$\fund$&$1$&$1$&$\fund$  \\
$\t q$&$\antifund$&$1$&$ \fund$ &$1$ \\
$u$&$\antifund$&$1$&$ 1$ &$1$ \\
$\t x$&$\antifund$&$\fund$&$ 1$ &$1$ \\
$a$&$ \antisym$&$1$&$1$ &$1$ \\
$R$&$1$&$1$&$\fund$ &$\antifund$ \\
$S$&$1$&$1$&$1$ &$\antifund$ \\
\end{tabular}
\ee
\end{center}
and the superpotential is
\be
W=\t q a \t q+q R \t q+q S u
+\t x a \t x \,.
\ee
This is a deformed version of the self-dual chiral theory presented in \cite{CEHT} and will be shown to be self-dual with the same methods. For now, let us focus on the case $F > 6$; we will present the dual for $F=5$ in \S\ref{subsec:fiveflavors},  and the dual for $F=6$ in \S\ref{subsec:sixflavors}. 

%%%%%%%%%%%%%%%%%%%%%%%%%%%%%%%%%%%%%%%%%%%
\subsubsection{Even $F$: A self-dual theory}

For $F> 6$ and even, a further dual description may be obtained by deconfining the antisymmetric tensor $a$. The resulting dual theory is given by another $SU(F-3)'$ gauge theory with matter

\begin{center}
\be  \label{tab:simpleir2}
\begin{tabular}{c|c|cccc}
&$SU(F-3)'$&$Sp(2F-8-F_1)$&$Sp(F_1)$&$SU(F)$   \\
\hline
&&&&&\\[-12pt]
$q_1$&$\fund$&$1$&$1$&$\antifund$  \\
$\t q_1$&$\antifund$&$1$&$ \antifund$ &$1$ \\
$u_1$&$\antifund$&$1$&$ 1$ &$1$ \\
$\t x_1$&$\antifund$&$\fund$&$ 1$ &$1$ \\
$a_1$&$ \antisym$&$1$&$1$ &$1$ \\
$r$&$1$&$\fund$&$1$ &$\fund$ \\
$s$&$1$&$1$&$1$ &$\fund$ \\
\end{tabular}
\ee
\end{center}
and the superpotential is
\be
W=\t q_1 a_1 \t q_1+ q r \t x_1 +q_1 s u_1
+\t x_1 a_1 \t x_1 \,.
\ee
Remarkably, {\it dualizing again by deconfinement returns the theory of} (\ref{tab:secondir}).\footnote{This is in contrast to previous examples found in the literature, characterized by an infinite tower of deconfinement steps and additional gauge groups. In our case, it is satisfying to find that the theory closes under further duality.} Note that this further duality step leaves the free $Sp(2F-8-F_1)$ sector essentially unchanged. It does, however, convert the irrelevant superpotential term $q l M \t x$ into an ostensibly marginal one, $l M r$.

Thus we see that  the dual description of the chiral SCFT consists of a free sector and a self-dual interacting sector in the far infrared. However, this duality was only apparent via deconfinement in the case of $N$ odd and $F$ even. It is natural to investigate whether the proposed duality generalizes straightforwardly to the various other possible parities of $N$ and $F$.

%%%%%%%%%%%%%%%%%%%%%%%%%%%%%%%%%%%%%%%%%%%
\subsubsection{Odd $F$: A dual with truncated global symmetries}

When $F$ is odd, the dual description to (\ref{tab:simpleir1}) is instead
\begin{center}
\be\label{table:simpleir2even}
\begin{tabular}{c|cc|ccc}
&$SU(F-2)$&$Sp(2F-8-F_1)$&$SU(2)$&$Sp(F_1)$&$SU(F)$   \\
\hline
&&&&&\\[-12pt]
$q_1$&$\fund$&$1$&$1$&$1$&$\antifund$  \\
$\t q_1$&$\antifund$&$1$&$1$&$ \fund$ &$1$ \\
$u_1$&$\antifund$&$1$&$\fund$&$1$ &$1$ \\
$\t x_1$&$\antifund$&$\fund$&$1$&$1$&$ 1$  \\
$a_1$&$ \antisym$&$1$&$ 1$&$1$&$1$  \\
$t$&$1$&$1$&$1$&$1$&$1$ \\
$r$&$1$&$\fund$&$1$&$1$ &$\fund$ \\
$s$&$1$&$1$&$\fund$&$1$ &$\fund$ \\
\end{tabular}
\ee
\end{center}
where now the superpotential is
\be \label{eq:Wevendual}
W=\t q_1 a_1 \t q_1+u_1 a_1 u_1 t+ q_1 \t x_1 r +  q_1 s u_1
+\t x_1 a_1 \t x_1 \,.
\ee

We see that this dual is qualitatively different from what we found in the $F$ even case. The ranks of the electric and magnetic theory do not agree; there are additional fields (beyond the expected meson $r = (q \t x)$) that are absent from the electric theory; and the global symmetries include an extra $SU(2)$. Let us discuss these points in more detail.

Regarding the rank and matter representations in the magnetic dual, the $SU(F-2)$ factor has a beta function coefficient equivalent to vector-like SQCD with $N_f=2N-1$.  So this factor is not at a self-dual point. However, this is crucial for the consistency of our proposal, because it allows us to close the duality circle. To see this, deconfine $a_1$ by introducing an $Sp(F-5)$ group. This does not require yet another global $SU(2)$ factor, and the $SU(F-2)$ group has $N_f=2(F-2)-1$ 
flavors.  Applying Seiberg duality to this node first gives an $SU(F-3)$ gauge group, which is precisely what is needed from the point of view of the original electric theory.  After integrating out heavy fields, we arrive at: 
\begin{center}
\be\label{table:simpleir2even2}
\begin{tabular}{c|cc|cccc}
&$SU(F-3)$&$Sp(F-5)$&$SU(2)$&$Sp(2F-8-F_1)$&$Sp(F_1)$&$SU(F)$  \\
\hline
&&&&\\[-12pt]
$q_2$&$\fund$&1&$1$&$1$&1&$ \fund$  \\
$\t q_2$&$\antifund$&1&$1$&1&$\fund$ &$1$ \\
$u_2$&$\antifund$&1&$\fund$&$ 1$ &$1$ & 1\\
$\t y$ & $\antifund$ & 1 & 1 & $\fund$ & 1 & 1 \\
$x_1'$&$\fund$&$\fund$&$ 1$&$ 1$ &$1$&1 \\
$u_1'$&$\antifund$&1&1&1&1&1\\
$(q_1 \t q_1)$&$1$&1&$ 1$&1&$ \fund$& $\antifund$  \\
$(q_1 u')$&1&1&1&1&1&$\antifund$ \\
$(x' u_1)$&$ 1$&$\fund$&$\fund$&$1$&1&1 \\
$t$&$1$&$1$&$1$ &$1$&1&1
\end{tabular}
\ee
\end{center}
with superpotential
\be\label{eq:Where}
W= \t q_2 x_1' x_1' \t q_2 + \t y x_1' x_1' \t y + (x'  u_1) (x' u_1) t + x_1' (x' u_1) u_2 + q_2 (q_1 u') u_1' + q_2 (q_1 \t q_1) \t q_2\,.
\ee
The $Sp(F-5)$ group s-confines, and after again integrating out heavy matter, 
we arrive at the original theory (\ref{tab:ir1}). This establishes the closure of dualities. 

The presence of additional fields and interactions -- together with the $SU(2)$ symmetry -- combine nontrivially to reproduce the moduli space of the electric theory. However, the $SU(2)$ symmetry itself is not a part of the quantum theory. All gauge invariants charged under the classical $SU(2)$ symmetry are eliminated from the chiral ring due to nonperturbative superpotentials (see below). Thus the $SU(2)$ symmetry does not exist at the quantum level, and the global symmetries of the second dual match those of (\ref{tab:simpleir1}). 

The same phenomenon was first observed in \cite{CEHT}. Although a detailed argument for truncation by nonperturbative effects was presented in \cite{CEHT}, let us review the argument for the specific theory considered here. Perhaps the simplest way to see the truncation of the $SU(2)$ symmetry is in (\ref{table:simpleir2even2}).  Let us study the effect of giving expectation values to the $SU(2)$-charged fields $u_2$ and $(x' u_1)$. First, a rank one expectation value for $u_2$ gives mass to two flavors of the $Sp$ gauge group; at low energies this factor confines and produces a constraint that breaks chiral symmetry. There is no simultaneous solution to this constraint and the F-term conditions from (\ref{eq:Where}), so a supersymmetric dual does not exist. Next, a rank-2 expectation value for $u_2$ reduces the effective number of flavors of the $Sp(F-5)$ theory to $F-5$. This theory has a nonperturbative superpotential~\cite{Intriligator:1995ne}
\be
W_\text{dyn} \sim \frac{\Lambda_{SP}^{F-2}}{(x_1')^{F-5}}
\ee
that yields a runaway to infinite field values. Thus a full-rank  $u_2$ is removed from the chiral ring of the supersymmetric theory. Identical arguments applied to $(x' u_1)$ yield a similar runaway. 

The important feature is that only singlets of the classical $SU(2)$ global symmetry remain once nonperturbative effects are taken into consideration. Thus we see that the additional global symmetries of the theory (\ref{table:simpleir2even}) are truncated quantum mechanically. Here deconfinement was a useful tool to determine the nonperturbative superpotential using existing techniques, but the nonperturbative superpotential may also be determined directly in the chiral theory using anomalous symmetries. Further explicit checks of this phenomenon may be made by studying the simplest theories with $F=5$ and $F=6$.

%%%%%%%%%%%%%%%%%%%%%%%%%%%%%%%%%%%%%%%%%%%
%%%%%%%%%%%%%%%%%%%%%%%%%%%%%%%%%%%%%%%%%%%
\subsection{Easy flavors from Seiberg duality}\label{subsec:easy}

Focusing on the interacting $SU(F-3)$ subsector, the cases $F=5$ and $F=6$ are interesting because our proposed dualities and nonperturbative effects may be checked directly using usual Seiberg duality without deconfinement.

%%%%%%%%%%%%%%%%%%%%%%%%%%%%%%%%%%%%%%%%%%%
\subsubsection{Five easy flavors}\label{subsec:fiveflavors}

Again setting aside the free fields charged only under $Sp(2F-8-F_1)$, the interacting sector for $F=5$ and $F_1=0$ is 
(see (\ref{tab:simpleir1}))
\begin{center}\be \label{tab:secondir}
\begin{tabular}{c|c|cc}
&$SU(2)$&$Sp(2)$&$SU(5)$  \\
\hline
&&&\\[-12pt]
$q$&$\fund$&1 &$\fund$  \\
$u$&$\fund$&$1$&$1$   \\
$\t x$&$\fund$& $\fund$ &$1$  \\
$a$&$1$ & $1$ &$1$  \\
$S$&$1$&$1$&$\antifund$ \\
\end{tabular}
\ee
\end{center}
with the customary superpotential
\be
W= q u S + \t x a \t x \,.
\ee
As an $SU(2)$ theory with 8 fundamentals, this theory is again self-dual: dualizing the $SU(2)$ yields 
\begin{center}\be \label{tab:secondir2}
\begin{tabular}{c|c|cc}
&$SU(2)'$&$Sp(2)$&$SU(5)$  \\
\hline
&&&\\[-12pt]
$q_1$&$\fund$&1 &$\antifund$  \\
$u_1$&$\fund$&$1$&$1$   \\
$\t x_1$&$\fund$& $\fund$ &$1$  \\
$(qq)$&$1$ & $1$ &$\antisym$  \\
$(q \t x)$&$1$&$\fund$&$\fund$ \\
$(u \t x)$ & $1$ & $\fund$ & $1$ \\
\end{tabular}
\ee
\end{center}
and superpotential
\be
W= q_1 (qq) q_1 + q_1 (q \t x) \t x_1 + u_1 (u \t x) \t x_1   \,.
\ee
Unsurprisingly, dualizing again returns the original $SU(2)$ theory with $(q_1 u_1) \sim S$ and $(\t x_1 \t x_1) \sim a$. Thus Seiberg duality alone suffices to show that the theory with $F=5$ is self-dual, and does not possess any additional quantum global symmetries. 

However, there is another possible dual. Using the proposed dual description (\ref{table:simpleir2even}) appropriate for odd $F$, the magnetic theory for $F=5$ should possess a dual description given by
\begin{center}
\be\label{table:theoryeven}
\begin{tabular}{c|c|ccc}
&$SU(3)$&$Sp(2)$&$SU(2)$&$SU(5)$   \\
\hline
&&&&\\[-12pt]
$q_1$&$\fund$&$1$&$1$&$\antifund$  \\
$u_1$&$\antifund$&$1$&$\fund$&$1$ \\
$\t x_1$&$\antifund$&$\fund$&$1$&$1$  \\
$a_1$&$ \antifund$&$1$&$ 1$&$1$  \\
$t$&$1$&$1$&$1$&$1$ \\
$r$&$1$&$\fund$&$1$&$\fund$ \\
$s$&$1$&$1$&$\fund$ &$\fund$ \\
\end{tabular}
\ee
\end{center}
where now the superpotential is
\be \label{eq:Wevendual2}
W=u_1 a_1 u_1 t+ q_1 x_1 r +  q_1 s u_1
+\t x_1 a_1 \t x_1 \,.
\ee
This theory has an extra $SU(2)$ global symmetry and additional degrees of freedom. 

Since the field $a_1$ is now an antifundamental flavor of $SU(3)$, we may dualize the $SU(3)$ factor using Seiberg duality for vector-like SQCD; this yields an $SU(2)$ theory with baryonic deformations. The baryonic deformations and superpotential terms give mass to a variety of fields; integrating them out leaves a theory with {\it no matter charged under the global} $SU(2)$. Rather, the remaining matter is
\begin{center}\be \label{tab:secondir3}
\begin{tabular}{c|c|cc}
&$SU(2)$&$Sp(2)$&$SU(5)$  \\
\hline
&&&\\[-12pt]
$q_2$&$\fund$&1 &$\fund$  \\
$a_2$&$\fund$&$1$&$1$   \\
$\t y$&$\fund$& $\fund$ &$1$  \\
$t$&$1$ & $1$ &$1$  \\
$(q_1a_1)$&$1$&$1$&$\antifund$ \\
\end{tabular}
\ee
\end{center}
with the customary superpotential
\be
W= q_2 (q_1a_1) a_2 + \t y t \t y \,.
\ee
After a relabeling of fields, this is precisely the original magnetic description with no additional $SU(2)$ global symmetry.

%%%%%%%%%%%%%%%%%%%%%%%%%%%%%%%%%%%%%%%%%%%
\subsubsection{Six easier flavors}\label{subsec:sixflavors}

In the case of six flavors, $F=6$, the dual gauge theory is $SU(3) \times Sp(4-F_1)$, and the antisymmetric tensor $a$ becomes merely another antifundamental flavor of $SU(3)$. The interacting $SU(3)$ sector is thus a vector-like self-dual theory with baryonic and singlet deformations.  The dual description is 

\begin{center}
\be  \label{tab:simpleir3}
\begin{tabular}{c|c|cccc}
&$SU(3)'$&$Sp(4-F_1)$&$Sp(F_1)$&$SU(6)$   \\
\hline
&&&&&\\[-12pt]
$q_1$&$\fund$&$1$&$1$&$\antifund$  \\
$\t q_1$&$\antifund$&$1$&$ \fund$ &$1$ \\
$u_1$&$\antifund$&$1$&$ 1$ &$1$ \\
$\t x_1$&$\antifund$&$\fund$&$ 1$ &$1$ \\
$a_1$&$ \antisym$&$1$&$1$ &$1$ \\
$(q \t x)$&$1$&$\fund$&$1$ &$\fund$ \\
$(q a)$&$1$&$1$&$1$ &$\fund$ \\
\end{tabular}
\ee
\end{center}
and superpotential after integrating out matter
\be
W=\t q_1 a_1 \t q_1+ q_1 (q \t x) \t x_1+q_1 (q a) u_1
+\t x_1 a_1 \t x_1 \,.
\ee
This agrees precisely with the dual description expected from a naive application of (\ref{tab:simpleir2}), though here it arises by Seiberg duality.  Dualizing again returns us to the original theory.

%%%%%%%%%%%%%%%%%%%%%%%%%%%%%%%%%%%%%%%%%%%
%%%%%%%%%%%%%%%%%%%%%%%%%%%%%%%%%%%%%%%%%%%
\subsection{Duality for even $N$ and an infinite family of dual descriptions}\label{subsec:even}

Let us now turn to the dual description of the chiral theory for even $N$. The duality map is similar to that of odd $N$, albeit somewhat more frequently populated with nonperturbatively-truncated global symmetries. In the case of $N$ odd, we saw that the dual magnetic description (\ref{tab:ir1}) possessed the same global symmetries as the electric description and contained an interacting sector that was either self-dual (for $F$ even) or dual to a related theory with an additional nonperturbatively-truncated classical symmetry (for $F$ odd). For $F$ of either parity, there was at least {\it one} magnetic description with {\it no} truncated classical symmetry, namely (\ref{tab:ir1}). 

One might be led to wonder whether the existence of another dual description with a nonperturbatively-truncated classical symmetry (as we found for $F$ odd) was simply an artifact of the particular choice of duality frame, and whether in general there is always a duality frame with no such truncated symmetry. We will see that is not the case; when $N$ is even and $F$ is odd we find that {\it all} possible dual descriptions exhibit a nonperturbatively-truncated classical symmetry. Thus the existence of such truncated symmetries seems to be an intrinsic property of the theory, and not simply a feature of particular duality frames.

%%%%%%%%%%%%%%%%%%%%%%%%%%%%%%%%%%%%%%%%%%%
\subsubsection{An infinite family of dual descriptions}

\begin{figure}[t] %  figure placement: here, top, bottom, or page
   \centering
   \includegraphics[width=4in]{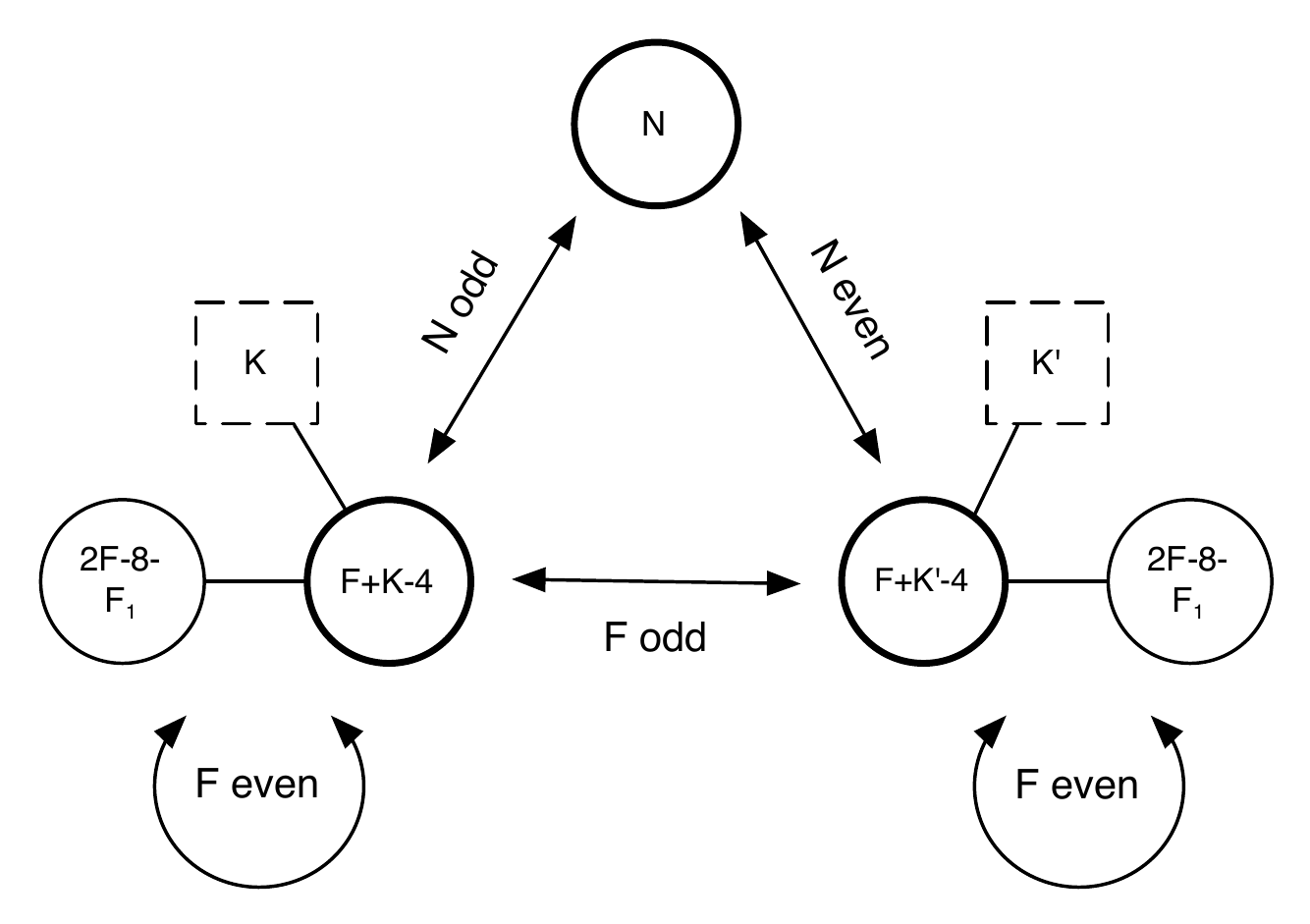} 
   \caption{A sketch of the general duality map. The chiral theory is dual to an infinite class of magnetic theories with classical global $SU(K), SU(K')$ symmetries that are truncated nonperturbatively. The duality relates all $SU(K)$ and $SU(K')$ theories such that $F+K+K'$ is even. All theories possess an additional $SU(F) \times SU(N+F-F_1-4)\times Sp(F_1)$ global symmetry that has been omitted for clarity.}
   \label{fig:gen}
\end{figure}

In its simplest form, the magnetic dual for even $N$ is a theory with gauge group $SU(F-2) \times Sp(2F-8-F_1)$, together with a global symmetry group given by the electric symmetries times an additional $SU(2)$ factor; see 
e.g.~Fig.~\ref{fig:map}. It turns out that this is a particular case of an \textit{infinite class} of duals characterized by arbitrarily large global symmetries and gauge groups, valid for $N$ of either parity. Rather than present the various duality maps for the theory with an $SU(2)$ symmetry, let us proceed directly to treat the most general case, with global symmetry $SU(K)$.  The full duality map is sketched in Fig.~\ref{fig:gen}.

These theories possess matter content
\begin{center}
\be \label{tab:ir1a}
\begin{tabular}{c|cc|ccccc}
&$SU(F+K-4)$&$Sp(2F-8-F_1)$&$SU(K)$&$SU(F_2)$&$Sp(F_1)$&$SU(F)$   \\
\hline
&&&&&\\[-12pt]
$q$&$\fund$&$1$&1&$1$&$1$&$\fund$  \\
$\t q$&$\antifund$&$1$&1&$1$&$ \fund$ &$1$ \\
$u$&$\antifund$&$1$&$\fund$&$1$&$ 1$ &$1$ \\
$\t x$&$\antifund$&$\fund$&$1$&$ 1$ &$1$&1 \\
$a$&$ \antisym$&$ 1$&$1$&$1$ &$1$&1 \\
$R$&$1$&$1$&1&$1$&$\fund$ &$\antifund$ \\
$S$&$1$&$1$&$\antifund$&$1$&$1$ &$\antifund$ \\
$T$ & 1&1&$\antiasymm$&1&1&1\\
\hline
&&&&\\[-12pt]
$l$&$1$&$\fund$&1&$\antifund$&$ 1$ &$1$ \\
$H$&$ 1$&$ 1$&1&$\antisym$&$1$ &$1$ \\
$M$&$1$&$1$&1&$\fund$&$1$ &$\antifund$ \\
\end{tabular}
\ee
\end{center}
and the superpotential is
\be\label{eq:Wir1a}
W=\t q a \t q+u a u T+q R \t q+q S u
+\t x a \t x +l H l+q M \t x l\,.
\ee
Here $K$ is such that $N+K$ is even (the special case $K = 1$ returns the $N$-odd theory of \S \ref{subsec:odd}, while $K=2$ returns the $N$-even theory of Fig.~\ref{fig:map}). These theories may be obtained from the electric theory by a generalized version of deconfinement. As we will show, the additional $SU(K)$ global symmetry is truncated by nonperturbative effects; all fields charged under $SU(K)$ are removed from the chiral ring. 

Under duality these theories flow to the same infrared fixed point as a second infinite class of theories enumerated by truncated $SU(K')$ global symmetries, with matter content
\begin{center}
\be \label{tab:ir3}
\begin{tabular}{c|cc|ccccc}
&$SU(F+K'-4)$&$Sp(2F-8-F_1)$&$SU(K')$&$SU(F_2)$&$Sp(F_1)$&$SU(F)$   \\
\hline
&&&&&\\[-12pt]
$q_1$&$\fund$&$1$&1&$1$&$1$&$\antifund$  \\
$\t q_1$&$\antifund$&$1$&1&$1$&$ \fund$ &$1$ \\
$u_1$&$\antifund$&$1$&$\antifund$&$1$&$ 1$ &$1$ \\
$\t x_1$&$\antifund$&$\fund$&$1$&$ 1$ &$1$&1 \\
$a_1$&$ \antisym$&$ 1$&$1$&$1$ &$1$&1 \\
$r$&$1$&$\fund$&1&$1$&$1$ &$\fund$ \\
$s$&$1$&$1$&$\fund$&$1$&$1$ &$\fund$ \\
$t$ & 1&1&$\antisym$&1&1&1\\
\hline
&&&&\\[-12pt]
$l$&$1$&$\fund$&1&$\antifund$&$ 1$ &$1$ \\
$H$&$ 1$&$ 1$&1&$\antisym$&$1$ &$1$ \\
$M$&$1$&$1$&1&$\fund$&$1$ &$\antifund$ \\
\end{tabular}
\ee
\end{center}
and the superpotential is
\be
W=\t q_1 a_1 \t q_1+u_1 a_1 u_1 t+q_1 r \t x_1 +q_1 s u_1
+\t x_1 a_1 \t x_1 +l H l + l M r\,.
\ee

Here $K'$ is any integer such that $F+K+K'$ is even. We propose that the infinite set of electric theories with fixed $F$ and arbitrary $K$ (provided $N$ and $K$ have the same parity)  is dual to the family of magnetic theories with the same $F$ and arbitrary $K'$ of the allowed parity. Note that for the theory with $N$ even, $F$ even, all possible dual descriptions possess an additional global symmetry truncated by nonperturbative effects. This suggests that the presence of a nonperturbatively-truncated global symmetry is not merely an artifact of the particular choice of duality frame, but instead arises in any magnetic duality frame.

Detailed tests of these dualities at the superconformal fixed point will be given in \S \ref{sec:amax}. In the rest of this section,  we examine the connection between $K$-dependent global and gauge symmetries, and the truncation of global symmetries by nonperturbative effects.

%%%%%%%%%%%%%%%%%%%%%%%%%%%%%%%%%%%%%%%%%%%
%%%%%%%%%%%%%%%%%%%%%%%%%%%%%%%%%%%%%%%%%%%
\subsubsection{$K$-dependent gauge and global symmetries}\label{subsec:gauge-global}

In the first sections, we found that an additional $SU(2)$ global symmetry was accompanied by an increase in the magnetic gauge group rank, from $SU(F-3)$ to $SU(F-2)$. This is made explicit in the general family of duals (\ref{tab:ir1a}), which features an $SU(K)$ global symmetry together with an $SU(F+K-4)$ magnetic gauge (sub)group.

It is then natural to ask how both phenomena are connected. It turns out that both are related by confinement effects and superpotential interactions. This may be illustrated using the product gauge group theory described in Appendix \ref{sec:decon}. The starting point is an electric theory with gauge group $SU(N) \times Sp(N+K-4)$ and global symmetries $SU(K) \times SU(F_2) \times Sp(F_1) \times SU(F)$; the matter content is given in (\ref{table:deconf-K}).

In this electric theory, the additional $SU(K)$ global group is naturally linked to the $K$ dependence in the gauge symmetry, such that the $Sp$ group s-confines. In more detail, when $\Lambda_{SU(N)} \gg \Lambda_{Sp(N+K-4)}$, the product gauge group theory flows to the magnetic dual (\ref{tab:ir1a}). However, we can also study strongly coupled effects by taking $\Lambda_{SU(N)} \ll \Lambda_{Sp(N+K-4)}$. In this case the $Sp$ group confines without breaking chiral symmetry. This gives a gapped theory without gauge interactions. In the confined theory, the superpotential interactions give masses to all matter charged under the global $SU(K)$. This establishes the connection between the $K$-dependent gauge and global symmetries via the product gauge group theory.

The disappearance of the $K$ dependence may also be understood directly in the magnetic theory. Next, we show how the classical symmetries are truncated by nonperturbative effects, and in \S \ref{sec:amax} we prove that the theory becomes independent of $K$ as it flows to the IR fixed point.

%%%%%%%%%%%%%%%%%%%%%%%%%%%%%%%%%%%%%%%%%%%
%%%%%%%%%%%%%%%%%%%%%%%%%%%%%%%%%%%%%%%%%%%
\subsubsection{Nonperturbative truncation of classical symmetries}\label{subsec:nonpert}

As we have seen, nonperturbative effects can truncate the chiral ring.  For the proposed infinite class of dualities considered here, the entire classical  $SU(K)$ and $SU(K')$ symmetries are removed quantum mechanically.
Consider giving $S$ a rank $n$ expectation value.  
For $n>1$, the following superpotential is generated:
\bea
\label{eq:NP2}
W_\text{dyn} =C_{N,K,n} \left(\frac{\Lambda_L^{F+2K+n-5}}{q^{F-n} a^{K-2} u^{K-n}}\right)^{1/(n-1)}\,,
\eea
where $C_{N,K,n}$ is a nonzero constant and 
$\Lambda_L^{F+2K+n-5}\sim \langle S^n \rangle \Lambda^{F+2K-5}$.  This dynamical superpotential leads to a runaway with no supersymmetric vacua, so $S$ is forced to have rank 1 or less.  For $S$ of rank 1 there is a quantum modified moduli space with supersymmetry breaking. We conclude that $S$ is not part of the chiral ring.

Now consider giving $T$ a rank $2n$ expectation value.  This gives rise to a nonperturbative superpotential of the form 
\be
\label{eq:NP3}
W_\text{dyn}=C'_{N,K,n} \left(\frac{\Lambda_L^{F+2K-5}}{q^{F}  a^{K-2-n} u^{K-2n}}\right)^{1/(n-1)}\,.
\ee
This leads to a runaway, 
removing $T$ from the quantum chiral ring.  The truncation of $s$ and $t$ from the $SU(K')$ theory proceeds in an analogous manner.
As discussed in \cite{CEHT}, it is possible to relate these nonperturbative effects to the familiar ADS superpotentials using deconfinement, which also allows one to explicitly show that $C_{N,K,n}$ and $C'_{N,K,n}$ are non-zero. 

All the gauge invariants that are charged under $SU(K)$ ($SU(K')$) are eliminated from the chiral ring of the magnetic $SU(F+K-4)$ ($SU(F+K'-4)$) theory. The classical flavor symmetries $SU(K)$, $SU(K')$ disappear nonperturbatively, and the magnetic global symmetry group that acts on the chiral ring is reduced to that of the original electric theory.

%%%%%%%%%%%%%%%%%%%%%%%%%%%%%%%%%%%%%%%%%%%
%%%%%%%%%%%%%%%%%%%%%%%%%%%%%%%%%%%%%%%%%%%
%%%%%%%%%%%%%%%%%%%%%%%%%%%%%%%%%%%%%%%%%%%
%%%%%%%%%%%%%%%%%%%%%%%%%%%%%%%%%%%%%%%%%%%
\section{Exact results from $a$-maximization}\label{sec:amax}

We have found that the chiral theory exhibits two novel features: (1) an infinite family of dual descriptions characterized by nonperturbatively-truncated global symmetries, and (2) an IR mixed phase (indeed, an infinite family of IR mixed phases) in which weakly-interacting fields and an IR-free gauge group are coupled to a conformal sector. Thus far we have provided a variety of evidence to support these claims, but even more concrete support may be obtained using $a$-maximization. In particular, we will use $a$-maximization to analyze all the phases of the theory that contain a conformal fixed point. These results lead to the phase diagram of Fig.~\ref{fig:phase} and establish, among other things, the IR freedom of the $Sp(2F-F_1-8)$ gauge group (thus confirming the existence of a mixed phase) and the independence of the dual descriptions on the classical $SU(K), SU(K')$ symmetries.

In a SCFT, the dimension of a gauge invariant operator, $\Delta_\OO$, is proportional to
its superconformal $R$ charge, $R_\OO$; for a spin zero field, the relation is $\Delta_{\mc O}= \frac{3}{2} R_{\mc O}$.
There are often many additional emergent $U(1)$ symmetries in the IR and it is not clear
which linear combination of $U(1)$ charges corresponds to the superconformal $R$-charge. 
In~\cite{Intriligator:2003jj}, it was shown that the superconformal $R$-charge can be determined
by maximizing the central charge
\be
\label{Eq:a-func}
a = \frac{3}{32} \left[ 3 \Tr R^3 - \Tr R \right]\,,
\ee
where the trace is done over all fermions in the theory.

The $a$-function is a measure of the number of degrees of freedom of the theory.
If the electric and magnetic theories are dual, they must describe the same physics in the far IR.  The propagating degrees of freedom at the fixed point should match.  
Matching the $R$ charges of chiral ring operators at the fixed point will then provide a very nontrivial test on our proposed dualities.\footnote{Note that matching the central charge $a$ itself is merely a consistency check, as agreement between the electric and magnetic descriptions is guaranteed by $U(1)$ anomaly matching. We thank Martin Schmaltz and 
Brian Wecht for helpful discussions of this point.}

The $R$ charges in the electric theory are obtained by maximizing (\ref{Eq:a-func}).
The precise $R$ charges obtained in this manner are somewhat complicated to list explicitly, but we will present various qualitative features. Particular conciseness may be obtained in the large-$N$ limit, allowing for the ratios $N/F$ and $N/F_1$ to be fixed. Here we present results in the large-$N$ limit, but have verified that corrections at small $N$ do not change the qualitative behavior.

%%%%%%%%%%%%%%%%%%%%%%%%%%%%%%%%%%%%%%%%%%%
%%%%%%%%%%%%%%%%%%%%%%%%%%%%%%%%%%%%%%%%%%%
\subsection{Chiral theory without superpotential}\label{subsec:amax1}

Let us begin with the electric theory with no superpotential, $F_1=0$. One finds that in the large $N$ limit, various gauge-invariant chiral operators go free as a function of $N$ and $F$. In particular, the gauge invariant $M = \t P Q$ goes free at $N=2.95367 F$.  To take into account $M$ going free, the $a$-function is modified to the form~\cite{Kutasov:2003iy,Barnes:2004jj}
\be
\label{Eq:a-func2}
a = \frac{3}{32} (\left[ 3 \Tr R^3 - \Tr R \right] - F_2 F \left[ 3 \Tr R_M^3 - \Tr R_M \right] + F_2 F \left[ 3 (-1/3)^3 -(-1/3) \right])
\ee
Proceeding again with $a$-maximization, one finds that in the large $N$ limit, the gauge invariant $H = \t Q A \t Q$ goes free at $N=4.08952 F$. Thus for $N \gtrsim 4 F$ both $M$ and $H$ are free fields in the chiral theory. 

Now let us apply $a$-maximization to the various dual descriptions, focusing on the more general theory (\ref{tab:ir1a}). The $R$ charges in the magnetic theory are obtained by maximizing (\ref{Eq:a-func})
subject to the constraints that the gauge groups $SU(F+K-4)$ and $Sp(2F-8-F_1)$ are anomaly free, and the superpotential (\ref{eq:Wir1a}) has $R$-charge 2.
Comparing the $R$ charges of the various gauge invariants, one finds that they match with those of the electric theory. Importantly, they are independent of the value of $K$, providing another check on the duality.

In the large $N$ limit, the field $M$ goes free at $N=2.95367 F$.  After $M$ goes free, the superpotential term, $M l q \t q$, becomes irrelevant.  The field $H$ subsequently goes free at the exact same value as it did in the electric theory. At this point, the whole $Sp$ factor becomes free, as we discuss in more detail below.

Throughout, the value of the $a$-functions match in the electric and magnetic theories, and both are independent of $K$.  An analogous set of  conclusions holds for the $SU(K')$ magnetic theory.\footnote{The conformal sectors of the magnetic duals are the same as those found in \cite{CEHT}, where more details of the $a$-maximization procedure are presented.} We point out that the case with no superpotential and $K=0$ or $2$ was studied using $a$-maximization in~\cite{Csaki:2004uj}. The results presented here are in agreement with their findings.

%%%%%%%%%%%%%%%%%%%%%%%%%%%%%%%%%%%%%%%%%%%
%%%%%%%%%%%%%%%%%%%%%%%%%%%%%%%%%%%%%%%%%%%
\subsection{Theory with superpotential interactions}\label{subsec:amax2}

The theory with nonzero superpotential $W_\text{el}=\t Q A \t Q$, can also be analyzed with $a$-maximization. Now we have to add the requirement that $W_\text{el}$ be marginal at the fixed point, namely, that $2=2 R_{\t Q}+ R_A$. 
The phase diagram now becomes two-dimensional, and is displayed in Fig.~\ref{fig:phase}. 

First, at small $x$ the results from both the electric and magnetic theory agree and imply the existence of a region in the phase diagram where the theory is at a nontrivial fixed point. All the gauge invariants have nonzero anomalous dimensions. This is the light grey region in Fig.~\ref{fig:phase}. Like before, $M=(Q \t P)$ is the first meson to become free; the electric and magnetic theory results agree after correcting for the fact that $M$ becomes free, and this corresponds to the red region. Next $H=(\t P A \t P)$ goes free, and the magnetic description implies that at this point the whole $Sp(2F-8-F_1)$ group goes free. This signals the appearance of the mixed phase --- the purple region in 
Fig.~\ref{fig:phase}. The results from the electric theory are no longer correct, because they do not capture the additional baryons that become free.

There is an exception to this: for $F_1 = 2 F -8$ the electric theory correctly captures the free $H$ meson. In this case, the $Sp$ gauge group ceases to exist and the gauge invariants of the electric and magnetic theories match even after $H$ goes free. This is the straight line in the figure, which divides the mixed phase from the white nonconformal region at large enough $y$. This has important implications for the existence of mixed phases. Indeed, it allows us to access both boundaries of the mixed phase region in Fig.~\ref{fig:phase} directly from the electric theory, thus proving that the proposed magnetic dual gives correct results. It also clarifies the physical origin of the mixed phase. Moving from the white to the purple region, the appearance of the mixed phase is similar to the enhancement of the s-confining SQCD to a full free magnetic gauge group. On the other hand, moving from the red region into the mixed phase, the interpretation is that of a theory that is so strongly coupled that even some baryons become free. Again, this is somewhat similar to the SQCD transition from the conformal window to the free magnetic range.

We note that certain positivity constraints (e.g.~$F_2>0$) in combination with $a$-maximization results imply interesting properties in the phase diagram; namely that for fixed $y=N/F_1$ and variable $x=N/F$, it is not always possible to explore all of the phases.  At large $F$ and $N$, requiring that $F_2 \ge 0$ implies $y^{-1}\le 1+ x^{-1}$. On the other hand, in the large $N$ limit and for 
$$
y  < \frac{205-\sqrt{73}}{456}\simeq 0.43\,,
$$ 
$M$ is never free. Indeed, for $y \lesssim 0.43$ the value of $x$ for which $M$ would go free does not satisfy the above bound. A similar statement holds for $H$, but for a larger value of $y$.

\begin{figure}[!t]
\centering
\includegraphics[width=4in]{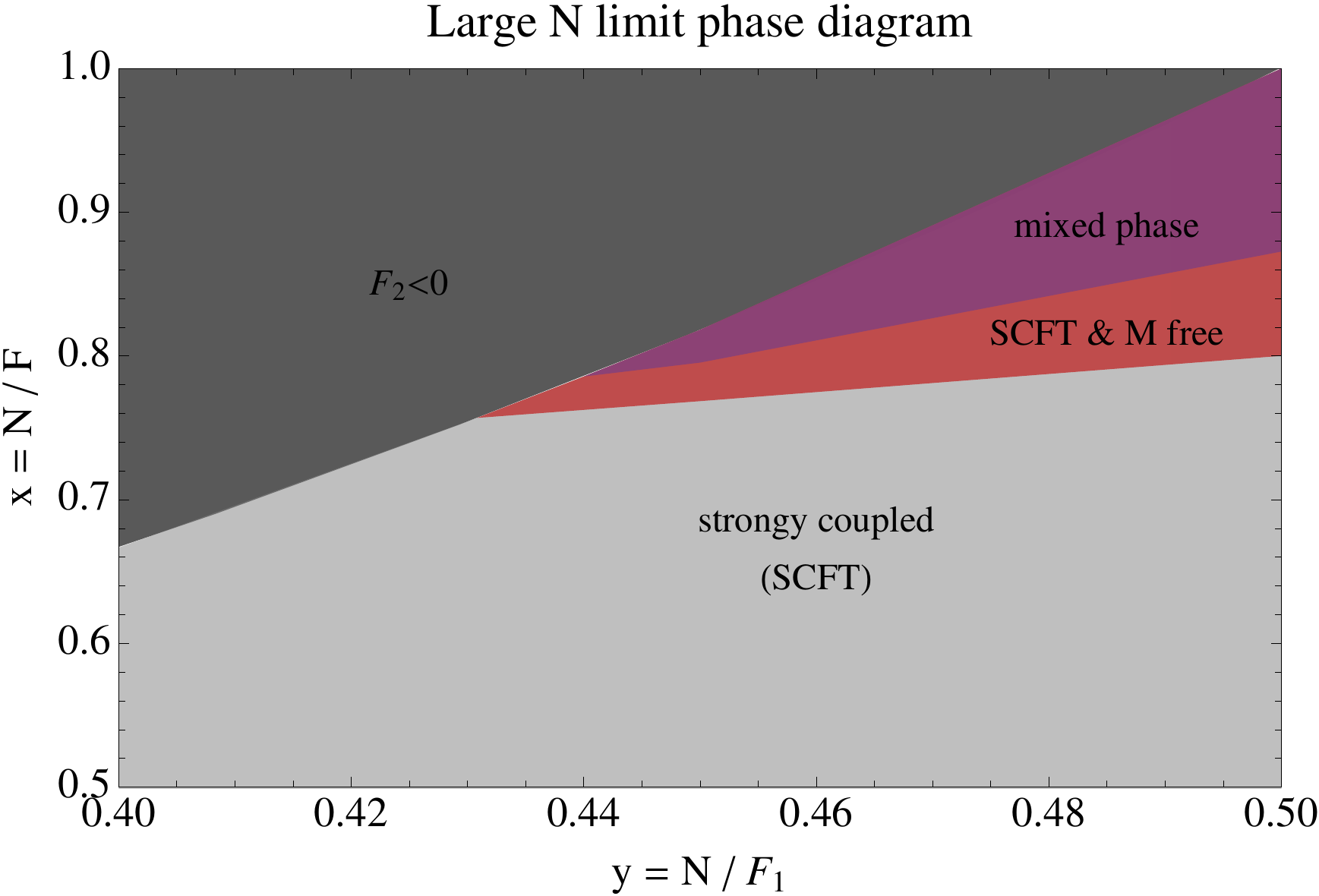}
\caption{Resolution of the `intersection point' between the different superconformal phases (same coloring 
scheme as in Fig.~1). $M$ and $H$ never become free simultaneously as a function of $(y,x)$. So there is no triple coexistence point between the light grey, red, and purple regions.}
\label{fig:phase2}
\end{figure}

Given the phase diagram in Fig.~\ref{fig:phase}, we need to understand whether there is a ``triple point'' where the three phases that contain a nontrivial fixed point meet. This would be the case if there existed a point for which, when $M$ goes free first (the boundary of the light grey region), $H$ also goes free. This could happen at a particular point on the boundary $y^{-1}= 1+ x^{-1}$. While finding the solution near this bound is complicated by numerical instabilities, we establish that $M$ and $H$ never become free simultaneously using the following strategy. Before $M$ goes free, (\ref{Eq:a-func}) for the electric theory should be maximized, and analytical answers can be obtained. After $M$ goes free, (\ref{Eq:a-func2}) should instead be used, making an analytic approach more involved. However, at the cross-over point when $M$ first becomes free, they both agree. The same considerations apply to $H$. Therefore, for the purpose of understanding whether $M$ and $H$ can become free simultaneously, it is enough to maximize
(\ref{Eq:a-func}) in the electric theory. We then find that $H$ never becomes free at the same time as $M$. So the intersection between the different phases is in fact resolved, giving rise to a phase diagram as depicted
in Fig.~\ref{fig:phase2}.

%%%%%%%%%%%%%%%%%%%%%%%%%%%%%%%%%%%%%%%%%%%
%%%%%%%%%%%%%%%%%%%%%%%%%%%%%%%%%%%%%%%%%%%
\subsection{Dynamics in the free magnetic subsector}\label{subsec:freemag}

Finally, let us explain in more detail why the gauge group $Sp(2F-8-F_1)$ is IR free after $H$ goes free.  This may be seen explicitly by going into the regime where $g_{Sp}$ is small (since the ratio $\Lambda_{Sp}/\Lambda_{SU}$ can be arbitrarily varied). The matter content of the general class of duals is given in (\ref{tab:ir1a}). Here the beta function for the $Sp(2F-8-F_1)$ theory is simply
\bea\label{eq:betagsp}
\beta(g_{Sp}) &=& -\frac{g_{Sp}^3}{16 \pi^2} \left[3 (2F-F_1-6) - (F+K-4)(1-\gamma_{\t x}(g_{Sp}=0))\right. \\ \nonumber
&-& \left.(F+N-4-F_1) (1-\gamma_l(g_{Sp}=0))\right] + \mathcal{O}(g_{Sp}^5)\,,
\eea
where the anomalous dimension $\gamma$ is related to the scaling dimension by $\Delta= 1 + \gamma/2$. We want to prove that when $H$ goes free first, the beta function vanishes to lowest order and that afterwards it changes sign. We do this self-consistently expanding around $g_{Sp}=0$.

The sign of the beta function is dictated by anomalous dimensions in the $g_{Sp}=0$ and $g_{SU} = \infty$ limit.  
The $R$-symmetry anomaly from the $Sp$ group is proportional to its gauge coupling constant, so once $g_{Sp}=0$ the constraint due to anomaly cancelation does not need to apply. Similar considerations apply to superpotential terms that become irrelevant when, for instance, certain gauge invariants become free fields. In this limit, the anomalous dimensions can be calculated with $a$-maximization using the assumption that the $Sp(2F-8-F_1)$ gauge group is a global symmetry.

The curve in $(y,x)$ space where $H$ goes free first defines the boundary between the red and purple regions in Fig.~\ref{fig:phase}. This occurs when $M$ has already gone free (as we also explained above). Therefore, the constraints on the $R$-charges are that the $SU(F+K-4)$ $R$-anomaly vanishes, and that all the terms in the superpotential (\ref{eq:Wir1a}) have $R$-charge $2$, except for the last one that is irrelevant. In particular, the requirement that $H$ goes free implies, from $W \supset l H l$, that the magnetic quarks $l$ also become free simultaneously.\footnote{Precisely at the cross-over where $H$ becomes free first, the superpotential interaction $W \supset l H l$ is still marginal, requiring $2=2R_l+R_H$. Once we move into the mixed phase region, this interaction becomes irrelevant.}

In this case, the beta function (\ref{eq:betagsp}) becomes
\be\label{eq:beta2}
\beta(g_{Sp}) \approx -\frac{g_{Sp}^3}{16 \pi^2} \frac{N}{x} \left(5-2\, \frac{x}{y}-x-\left(1+ \frac{K}{N}x \right)(1- \gamma_{\tilde x}) \right)\,,
\ee
where for simplicity we took a large $N$ limit, keeping $x$, $y$ and $K/N$ fixed. The anomalous dimension also depends on $K$, in such a way that the $K$ dependence drops out from the beta function. Solving numerically for the $a$-maximization conditions reveals that the locus where $H$ becomes free first indeed corresponds to the vanishing of (\ref{eq:beta2}). Starting from this curve, an increase in $x$ or a decrease in $y$ give rise to a positive beta function, thus establishing the IR freedom of the $Sp(2F-8-F_1)$ gauge group.

While we have not found an analytic expression for $\gamma_{\tilde x}$ at arbitrary $F_1$, the above results can be illustrated very concretely when $F_1=0$. At large $N$ we find that
\be
\beta(g_{Sp}) \approx -\frac{g_{Sp}^3}{16 \pi^2}  \frac{4.08952-x}{x} N\,.
\ee
At the exact value $x=4.08952$ when $H$ becomes a free field, the beta function switches signs.  Therefore, $H$ becoming free triggers the onset of a free magnetic phase in the gauge group $Sp(2F-8)$.

%%%%%%%%%%%%%%%%%%%%%%%%%%%%%%%%%%%%%%%%%%%
%%%%%%%%%%%%%%%%%%%%%%%%%%%%%%%%%%%%%%%%%%%
%%%%%%%%%%%%%%%%%%%%%%%%%%%%%%%%%%%%%%%%%%%
%%%%%%%%%%%%%%%%%%%%%%%%%%%%%%%%%%%%%%%%%%%
\section{Discussion}\label{sec:discussion}

In this work, we have studied the IR dynamics of supersymmetric chiral gauge theories with an antisymmetric tensor. The presence of a marginal superpotential interaction provides a probe for exploring different phases of the theory, and makes manifest various types of dynamical effects and phase transitions. The phase diagram, which at large $N$ is two dimensional, is shown in Fig.~\ref{fig:phase}. Various regions of the phase diagram may be described by new dual descriptions presented in this work. Such dual descriptions allow us to understand, among other things, the appearance and properties of mixed phases. We believe our results put the existence of such mixed phases in $\mc N=1$ supersymmetric gauge theories on a firmer footing.

There are various interesting directions for future research. At a more formal and general level, we hope that our results motivate further developments on dualities for chiral gauge theories. We have seen how a combination of deconfinement, holomorphy/symmetry arguments, and $a$-maximization give us powerful handles on the long distance properties of these theories. Many other classes of chiral theories may be studied in a similar fashion. Moreover, it would be very interesting to apply the results of \cite{index} on superconformal indices to larger classes of chiral gauge theories. This may be a valuable tool for exploring such theories systematically.

The existence of mixed phases is an intriguing phenomenon, and it would be interesting to find other examples. Thus far, we have only found evidence for  $\mc N=1$ mixed phases in chiral gauge theories; it is natural to wonder whether there are vector-like $\mc N=1$ theories that exhibit the same phenomenon. Intriguingly, mixed phases have already been discovered in the context of $\mc N=2$ theories (which are inherently vector-like) by Argyres and Seiberg \cite{Argyres:2007cn}. Concretely, one could try to find calculable flows of such theories from 8 to 4 supercharges (perhaps along the lines of \cite{Benini:2009mz}), and further study the dynamics that emerges.

We also believe that our results may have many useful phenomenological applications, some of which we hope to explore in a future work. These chiral theories provide a clean, controlled environment in which one can study interactions between a nontrivial SCFT and a weakly coupled theory. In this respect, they provide a natural realization of ``hidden valley'' or ``quirky'' scenarios of great phenomenological interest \cite{hiddenvalleys}. It is also possible that these theories may be connected to our visible sector. In the UV they have the attractive feature of comprising merely a single sector, while in the IR they flow to a rich set of free and interacting fields. We envision applications to supersymmetry breaking and Higgs physics, as well as generalizations of technicolor with supersymmetry, perhaps along the lines of \cite{applications}.

%%%%%%%%%%%%%%%%%%%%%%%%%%%%%%%%%%%%%%%%%%%%%%%%%%%%%%%%%%
%%%%%%%%%%%%%%%%%%%%%%%%%%%%%%%%%%%%%%%%%%%%%%%%%%%%%%%%%%
\acknowledgments
We thank Shamit Kachru, Martin Schmaltz, John Terning, and Brian Wecht for useful discussions. NC is supported in part by the National Science Foundation under Grant No.s~PHY09-07744 and gratefully acknowledges support from the Institute for Advanced Study.
AH, RE, and GT are supported by the US DOE under contract number DE-AC02-76SF00515. RE also acknowledges support by the National Science Foundation under Grant No.~PHY-0969739 and PHY05-51164.

%%%%%%%%%%%%%%%%%%%%%%%%%%%%%%%%%%%%%%%%%%%%%%%%%%%%%%%%%%
%%%%%%%%%%%%%%%%%%%%%%%%%%%%%%%%%%%%%%%%%%%%%%%%%%%%%%%%%%
\appendix

%%%%%%%%%%%%%%%%%%%%%%%%%%%%%%%%%%%%%%%%%%%
%%%%%%%%%%%%%%%%%%%%%%%%%%%%%%%%%%%%%%%%%%%
%%%%%%%%%%%%%%%%%%%%%%%%%%%%%%%%%%%%%%%%%%%
%%%%%%%%%%%%%%%%%%%%%%%%%%%%%%%%%%%%%%%%%%%
\section{Product gauge group theory and deconfinement}\label{sec:decon}

In this Appendix we explain the deconfinement procedure that allows us to determine appropriate dual descriptions. 
Consider the electric theory with $F_1$ flavors of $\t Q$ coupled to the antisymmetric $A$. The theory is then given by
\begin{center}
\be\label{table:chiral1}
\begin{tabular}{c|c|ccc}
&$SU(N)$&$SU(F_2)$&$Sp(F_1)$&$SU(F)$  \\
\hline
&&&&\\[-12pt]
$Q$&$\fund$&$1$&$1$&$\antifund$  \\
$\t Q$&$\antifund$&$1$&$ \fund$ &$1$ \\
$\t P$&$\antifund$&$\fund$&$1$&$1$  \\
$A$&$\antisym$&$1$&$1$&$1$  \\
\end{tabular}
\ee
\end{center}
and the superpotential is
\be\label{eq:Wchiral2}
W_\text{el}=  \t Q A \t Q\,.
\ee
We have defined the combination
\be
F_2 \equiv N+F-F_1-4\,.
\ee

The basic idea is to find a product gauge group theory such that when the new factor confines, (\ref{table:chiral1}) is reproduced. New dual descriptions are then obtained by inverting the order of the dynamical scales and dualizing the $SU(N)$ factor first.

In detail, the product gauge group theory is
\begin{center}
\be\label{table:deconf-K}
\begin{tabular}{c|cc|cccc}
&$SU(N)$&$Sp(N+K-4)$&$SU(K)$&$SU(F_2)$&$Sp(F_1)$&$SU(F)$  \\
\hline
&&&&&\\[-12pt]
$Q$&$\fund$&$1$&$1$&$1$&$1$&$\antifund$  \\
$\t Q$&$\antifund$&$1$&$1$&$1$&$ \fund$ &$1$ \\
$\t P$&$\antifund$&$1$&$1$&$\fund$&$1$&$1$  \\
$X$&$ \fund$&$\fund$&$ 1$&$1$&$1$ &$1$ \\
$U$&$\antifund$&$1$&$\antifund$&$1$&$ 1$ &$1$ \\
$V$&$1$&$\fund$&$\fund$&$ 1$&$1$ &$1$ \\
$T$&$1$&$1$&$\overline\antisym$&$1$&$1$ &$1$
\end{tabular}
\ee
\end{center}
The electric superpotential is
\be\label{eq:Wdeconf}
W=  \t Q XX \t Q+XUV+VVT\,.
\ee
The appearance of additional global symmetries was first pointed out in~\cite{Luty:1996cg}.

In the regime $\Lambda_{Sp(N+K-4)} \gg \Lambda_{SU(N)}$, the $Sp$ factor becomes strong first and s-confines. This generates an antisymmetric meson $(XX)$ that is identified with $A$, plus the mesons $(XV)$ and $(VV)$. The additional fields $U$ and $T$, together with the superpotential (\ref{eq:Wdeconf}) ensure that, after confinement, the theory 
(\ref{table:chiral1}) is recovered.

In order to derive a new magnetic description, we invert the order of scales, taking $\Lambda_{Sp(N+K-4)} \ll \Lambda_{SU(N)}$. Then we dualize $SU(N)$ first, obtaining
\begin{center}
\begin{tabular}{c|cc|cccc}
&$SU(F+K-4)$&$Sp(N+K-4)$&$SU(K)$&$SU(F_2)$&$Sp(F_1)$&$SU(F)$   \\
\hline
&&&&&\\[-12pt]
$q$&$\fund$&$1$&$1$&$1$&$1$&$\fund$  \\
$\t q$&$\antifund$&$1$&$1$&$1$&$ \fund$ &$1$ \\
$\t p$&$\antifund$&$1$&$1$&$\antifund$&$1$&$1$  \\
$x$&$ \fund$&$\fund$&$ 1$&$1$&$1$ &$1$ \\
$u$&$\antifund$&$1$&$\fund$&$1$&$ 1$ &$1$ \\
$V$&$1$&$\fund$&$\fund$&$1$&$ 1$ &$1$ \\
$T$&$1$&$1$&$\overline\antisym$&$1$&$1$ &$1$ \\
$(Q \t Q)$&$1$&$1$&$1$&$1$&$\fund$ &$\antifund$ \\
$(Q \t P)$&$1$&$1$&$1$&$\fund$&$1$ &$\antifund$ \\
$(Q  U)$&$1$&$1$&$\antifund$&$1$&$1$ &$\fund$ \\
$(X \t P)$&$1$&$\fund$&$1$&$\fund$&$1$ &$1$ \\
$(X \t Q)$&$1$&$\fund$&$1$&$1$&$\fund$ &$1$ \\
$(X U)$&$1$&$\fund$&$\antifund$&$1$&$1$ &$1$
\end{tabular}
\end{center}
The superpotential now reads
\bea
W&=&  (\t Q X) (\t Q X)+(XU)V+VVT+\nonumber\\
&+&q (Q \t Q) \t q+q (Q \t P) \t p+q (Q U) u+x (X \t P) \t p+x (X \t Q) \t q+x (X U) u\,.
\eea
Importantly, in this duality frame the original electric superpotential $W_\text{el}= \t Q A \t Q$ produces mass terms $W \supset (\t QX)^2$. We will see shortly that this is responsible for the $F_1$ dependence in the rank of the free magnetic subsector.

Integrating out the massive fields leads to
\begin{center}
\be\label{table:interm}
\begin{tabular}{c|cc|cccc}
&$SU(F+K-4)$&$Sp(N+K-4)$&$SU(K)$&$SU(F_2)$&$Sp(F_1)$&$SU(F)$   \\
\hline
&&&&&\\[-12pt]
$q$&$\fund$&$1$&$1$&$1$&$1$&$\fund$  \\
$\t q$&$\antifund$&$1$&$1$&$1$&$ \fund$ &$1$ \\
$\t p$&$\antifund$&$1$&$1$&$\antifund$&$1$&$1$  \\
$x$&$ \fund$&$\fund$&$ 1$&$1$&$1$ &$1$ \\
$u$&$\antifund$&$1$&$\fund$&$1$&$ 1$ &$1$ \\
$T$&$1$&$1$&$\overline\antisym$&$1$&$1$ &$1$ \\
$(Q \t Q)$&$1$&$1$&$1$&$1$&$\fund$ &$\antifund$ \\
$(Q \t P)$&$1$&$1$&$1$&$\fund$&$1$ &$\antifund$ \\
$(Q  U)$&$1$&$1$&$\antifund$&$1$&$1$ &$\fund$ \\
$(X \t P)$&$1$&$\fund$&$1$&$\fund$&$1$ &$1$ \\
\end{tabular}
\ee
\end{center}
and
\be\label{eq:Winterm}
W= \t q xx \t q+xuxuT+q (Q \t Q) \t q+q (Q \t P) \t p+q (Q U) u+x (X \t P) \t p\,.
\ee

Dualizing the $Sp$ gives
\begin{center}
\begin{tabular}{c|cc|cccc}
&$SU(F+K-4)$&$Sp(2F-8-F_1)$&$SU(K)$&$SU(F_2)$&$Sp(F_1)$&$SU(F)$   \\
\hline
&&&&&\\[-12pt]
$q$&$\fund$&$1$&$1$&$1$&$1$&$\fund$  \\
$\t q$&$\antifund$&$1$&$1$&$1$&$ \fund$ &$1$ \\
$\t p$&$\antifund$&$1$&$1$&$\antifund$&$1$&$1$  \\
$u$&$\antifund$&$1$&$\fund$&$1$&$ 1$ &$1$ \\
$\t x$&$\antifund$&$\fund$&$1$&$1$&$ 1$ &$1$ \\
$r$&$1$&$\fund$&$1$&$\antifund$&$ 1$ &$1$ \\
$(xx)$&$ \antisym$&$1$&$ 1$&$1$&$1$ &$1$ \\
$(x(X\t P))$&$ \fund$&$1$&$ 1$&$\fund$&$1$ &$1$ \\
$((X\t P)^2)$&$ 1$&$1$&$ 1$&$\antisym$&$1$ &$1$ \\
$T$&$1$&$1$&$\antiasymm$&$1$&$1$ &$1$ \\
$(Q \t Q)$&$1$&$1$&$1$&$1$&$\fund$ &$\antifund$ \\
$(Q \t P)$&$1$&$1$&$1$&$\fund$&$1$ &$\antifund$ \\
$(Q  U)$&$1$&$1$&$\antifund$&$1$&$1$ &$\fund$ \\
\end{tabular}
\end{center}
where $\t x$ and $r$ are dual to $x$ and $(X \t P)$, respectively. The superpotential reads
\bea
W&=& \t q (xx) \t q+u(xx)uT+q (Q \t Q) \t q+q (Q \t P) \t p+q (Q U) u+(x (X \t P)) \t p+\nonumber\\
&+&\t x (xx) \t x + \t x (x(X \t P))r+r((X\t P)^2) r\,.
\eea

Now the meson $(x (X \t P))$ acquires a mass term with $\t p$. Integrating it out, we obtain the low energy theory
\begin{center}
\begin{tabular}{c|cc|cccc}
&$SU(F+K-4)$&$Sp(2F-8-F_1)$&$SU(K)$&$SU(F_2)$&$Sp(F_1)$&$SU(F)$   \\
\hline
&&&&&\\[-12pt]
$q$&$\fund$&$1$&$1$&$1$&$1$&$\fund$  \\
$\t q$&$\antifund$&$1$&$1$&$1$&$ \fund$ &$1$ \\
$u$&$\antifund$&$1$&$\fund$&$1$&$ 1$ &$1$ \\
$\t x$&$\antifund$&$\fund$&$1$&$1$&$ 1$ &$1$ \\
$r$&$1$&$\fund$&$1$&$\antifund$&$ 1$ &$1$ \\
$(xx)$&$ \antisym$&$1$&$ 1$&$1$&$1$ &$1$ \\
$((X\t P)^2)$&$ 1$&$1$&$ 1$&$\antisym$&$1$ &$1$ \\
$T$&$1$&$1$&$\antiasymm$&$1$&$1$ &$1$ \\
$(Q \t Q)$&$1$&$1$&$1$&$1$&$\fund$ &$\antifund$ \\
$(Q \t P)$&$1$&$1$&$1$&$\fund$&$1$ &$\antifund$ \\
$(Q  U)$&$1$&$1$&$\antifund$&$1$&$1$ &$\fund$ \\
\end{tabular}
\end{center}
and the superpotential is
\be
W=\t q (xx) \t q+u(xx)uT+q (Q \t Q) \t q+q (Q \t P) \t x r+q (Q U) u
+\t x (xx) \t x +r((X\t P)^2) r\,.
\ee

After a renaming of fields, this reproduces the dual presented in (\ref{tab:ir1a}). A further application of deconfinement leads to (\ref{tab:ir3}).

%%%%%%%%%%%%%%%%%%%%%%%%%%%%%%%%%%%%%%%%%%%
%%%%%%%%%%%%%%%%%%%%%%%%%%%%%%%%%%%%%%%%%%%
%%%%%%%%%%%%%%%%%%%%%%%%%%%%%%%%%%%%%%%%%%%
%%%%%%%%%%%%%%%%%%%%%%%%%%%%%%%%%%%%%%%%%%%


\begin{thebibliography}{100}


\bibitem{Affleck:1984uz}
  I.~Affleck, M.~Dine and N.~Seiberg,
  %``Calculable Nonperturbative Supersymmetry Breaking,''
  Phys.\ Rev.\ Lett.\  {\bf 52}, 1677 (1984).
  %%CITATION = PRLTA,52,1677;%%

%\cite{Affleck:1984xz}
\bibitem{Affleck:1984xz}
  I.~Affleck, M.~Dine and N.~Seiberg,
  %``Dynamical Supersymmetry Breaking In Four-Dimensions And Its
  %Phenomenological Implications,''
  Nucl.\ Phys.\  B {\bf 256}, 557 (1985).
  %%CITATION = NUPHA,B256,557;%%


%\cite{Pouliot:1995zc}
\bibitem{Pouliot:1995zc}
  P.~Pouliot,
  %``Chiral duals of nonchiral SUSY gauge theories,''
  Phys.\ Lett.\  {\bf B359}, 108-113 (1995).
  [hep-th/9507018].

%\cite{Pouliot:1996zh}
\bibitem{Pouliot:1996zh}
  P.~Pouliot, M.~J.~Strassler,
  %``Duality and dynamical supersymmetry breaking in Spin(10) with a spinor,''
  Phys.\ Lett.\  {\bf B375}, 175-180 (1996).
  [hep-th/9602031].
 P.~Pouliot, M.~J.~Strassler,
  %``A Chiral SU(n) gauge theory and its nonchiral spin(8) dual,''
  Phys.\ Lett.\  {\bf B370}, 76-82 (1996).
  [hep-th/9510228].
  
\bibitem{Pouliot:1995me}
  P.~Pouliot,
  %``Duality in SUSY $SU(N)$ with an Antisymmetric Tensor,''
  Phys.\ Lett.\  B {\bf 367}, 151 (1996)
  [arXiv:hep-th/9510148].
  %%CITATION = PHLTA,B367,151;%%


\bibitem{Terning:1997jj}
  J.~Terning,
  %``Duals for SU(N) SUSY gauge theories with an antisymmetric tensor: Five
  %easy flavors,''
  Phys.\ Lett.\  B {\bf 422}, 149 (1998)
  [arXiv:hep-th/9712167].
  %%CITATION = PHLTA,B422,149;%%

%\cite{Csaki:2004uj}
\bibitem{Csaki:2004uj}
  C.~Csaki, P.~Meade and J.~Terning,
  %``A mixed phase of SUSY gauge theories from a-maximization,''
  JHEP {\bf 0404}, 040 (2004)
  [arXiv:hep-th/0403062].
  %%CITATION = JHEPA,0404,040;%%

%\cite{Intriligator:1995ax}
\bibitem{Intriligator:1995ax}
  K.~A.~Intriligator, R.~G.~Leigh, M.~J.~Strassler,
  %``New examples of duality in chiral and nonchiral supersymmetric gauge theories,''
  Nucl.\ Phys.\  {\bf B456}, 567-621 (1995).
  [hep-th/9506148].

%\cite{Brodie:1996xm}
\bibitem{Brodie:1996xm}
  J.~H.~Brodie, M.~J.~Strassler,
  %``Patterns of duality in N=1 SUSY gauge theories, or: Seating preferences of theater going nonAbelian dualities,''
  Nucl.\ Phys.\  {\bf B524}, 224-250 (1998).
  [hep-th/9611197].
  
%\cite{Seiberg:1994bz}
\bibitem{Seiberg:1994bz}
  N.~Seiberg,
  %``Exact results on the space of vacua of four-dimensional SUSY gauge theories,''
  Phys.\ Rev.\  {\bf D49}, 6857-6863 (1994).
  [hep-th/9402044].
N.~Seiberg,
  %``Electric - magnetic duality in supersymmetric nonAbelian gauge theories,''
  Nucl.\ Phys.\  {\bf B435}, 129-146 (1995).
  [hep-th/9411149].

%\cite{Intriligator:1995au}
\bibitem{Intriligator:1995au}
  K.~A.~Intriligator, N.~Seiberg,
  %``Lectures on supersymmetric gauge theories and electric - magnetic duality,''
  Nucl.\ Phys.\ Proc.\ Suppl.\  {\bf 45BC}, 1-28 (1996).
  [hep-th/9509066].
  
  %\cite{Berkooz:1995km}
\bibitem{Berkooz:1995km}
  M.~Berkooz,
  %``The Dual of supersymmetric SU(2k) with an antisymmetric tensor and
  %composite dualities,''
  Nucl.\ Phys.\  B {\bf 452}, 513 (1995)
  [arXiv:hep-th/9505067].
  %%CITATION = NUPHA,B452,513;%%

%\cite{CEHT}
\bibitem{CEHT}
 N.~Craig, R.~Essig, A.~Hook and G.~Torroba,
  %``New dynamics and dualities in supersymmetric chiral gauge theories,''
  JHEP {\bf 1109}, 046 (2011)
  [arXiv:1106.5051 [hep-th]].
  %%CITATION = JHEPA,1109,046;%%


%\cite{Seiberg:1994rs}
%\bibitem{Seiberg:1994rs}
%  N.~Seiberg, E.~Witten,
  %``Electric - magnetic duality, monopole condensation, and confinement in N=2 supersymmetric Yang-Mills theory,''
%  Nucl.\ Phys.\  {\bf B426}, 19-52 (1994).
%  [hep-th/9407087].
  


%\cite{Intriligator:2003jj}
\bibitem{Intriligator:2003jj}
  K.~A.~Intriligator and B.~Wecht,
  %``The exact superconformal R-symmetry maximizes a,''
  Nucl.\ Phys.\  B {\bf 667}, 183 (2003)
  [arXiv:hep-th/0304128].
  %%CITATION = NUPHA,B667,183;%%

\bibitem{future}
N.~Craig, R.~Essig, A.~Hook and G.~Torroba, to appear.
  
\bibitem{Intriligator:1995ne}
  K.~A.~Intriligator and P.~Pouliot,
  %``Exact superpotentials, quantum vacua and duality in supersymmetric SP(N(c))
  %gauge theories,''
  Phys.\ Lett.\  B {\bf 353}, 471 (1995)
  [arXiv:hep-th/9505006].
  %%CITATION = PHLTA,B353,471;%%

\bibitem{Luty:1996cg}
  M.~A.~Luty, M.~Schmaltz and J.~Terning,
%  ``A Sequence of Duals for Sp(2N) Supersymmetric Gauge Theories with Adjoint
%  Matter,''
  Phys.\ Rev.\  D {\bf 54}, 7815 (1996)
  [arXiv:hep-th/9603034].
  %%CITATION = PHRVA,D54,7815;%%



%\cite{Kutasov:2003iy}
\bibitem{Kutasov:2003iy}
  D.~Kutasov, A.~Parnachev and D.~A.~Sahakyan,
  %``Central charges and U(1)R symmetries in N = 1 super Yang-Mills,''
  JHEP {\bf 0311}, 013 (2003)
  [arXiv:hep-th/0308071].
  %%CITATION = JHEPA,0311,013;%%

%\cite{Barnes:2004jj}
\bibitem{Barnes:2004jj}
  E.~Barnes, K.~A.~Intriligator, B.~Wecht and J.~Wright,
  %``Evidence for the strongest version of the 4d a-theorem, via a-maximization
  %along RG flows,''
  Nucl.\ Phys.\  B {\bf 702}, 131 (2004)
  [arXiv:hep-th/0408156].
  %%CITATION = NUPHA,B702,131;%%

\bibitem{index}
C.~Romelsberger,
  %``Calculating the Superconformal Index and Seiberg Duality,''
  [arXiv:0707.3702 [hep-th]].

C.~Romelsberger,
  %``Counting chiral primaries in N = 1, d=4 superconformal field theories,''
  Nucl.\ Phys.\  {\bf B747}, 329-353 (2006).
  [hep-th/0510060].
  
F.~A.~Dolan, H.~Osborn,
  %``Applications of the Superconformal Index for Protected Operators and q-Hypergeometric Identities to N=1 Dual Theories,''
  Nucl.\ Phys.\  {\bf B818}, 137-178 (2009).
  [arXiv:0801.4947 [hep-th]].

V.~P.~Spiridonov, G.~S.~Vartanov,
  %``Elliptic hypergeometry of supersymmetric dualities,''
  Commun.\ Math.\ Phys.\  {\bf 304}, 797-874 (2011).
  [arXiv:0910.5944 [hep-th]].
  
V.~P.~Spiridonov, G.~S.~Vartanov,
  %``Elliptic hypergeometry of supersymmetric dualities II. Orthogonal groups, knots, and vortices,''
  [arXiv:1107.5788 [hep-th]].

%\cite{Argyres:2007cn}
\bibitem{Argyres:2007cn}
  P.~C.~Argyres, N.~Seiberg,
  %``S-duality in N=2 supersymmetric gauge theories,''
  JHEP {\bf 0712}, 088 (2007).
  [arXiv:0711.0054 [hep-th]].
  
  \bibitem{Benini:2009mz}
  F.~Benini, Y.~Tachikawa, B.~Wecht,
  %``Sicilian gauge theories and N=1 dualities,''
  JHEP {\bf 1001}, 088 (2010).
  [arXiv:0909.1327 [hep-th]].

\bibitem{hiddenvalleys}

  M.~J.~Strassler, K.~M.~Zurek,
  %``Echoes of a hidden valley at hadron colliders,''
  Phys.\ Lett.\  {\bf B651}, 374-379 (2007).
  [hep-ph/0604261]
  
  J.~Kang, M.~A.~Luty,
  %``Macroscopic Strings and 'Quirks' at Colliders,''
  JHEP {\bf 0911}, 065 (2009).
  [arXiv:0805.4642 [hep-ph]].
  
  C.~Kilic, T.~Okui, R.~Sundrum,
  %``Vectorlike Confinement at the LHC,''
  JHEP {\bf 1002}, 018 (2010).
  [arXiv:0906.0577 [hep-ph]].
  
  

\bibitem{applications}
  N.~Craig, D.~Stolarski, J.~Thaler,
  %``A Fat Higgs with a Magnetic Personality,''  
  [arXiv:1106.2164 [hep-ph]].

A.~Azatov, J.~Galloway, M.~A.~Luty,
  %``Superconformal Technicolor: Models and Phenomenology,''
  [arXiv:1106.4815 [hep-ph]].

T.~Gherghetta, A.~Pomarol,
  %``A Distorted MSSM Higgs Sector from Low-Scale Strong Dynamics,''
  [arXiv:1107.4697 [hep-ph]].

  J.~J.~Heckman, P.~Kumar, C.~Vafa, B.~Wecht,
  %``Electroweak Symmetry Breaking in the DSSM,''
    [arXiv:1108.3849 [hep-ph]].

\end{thebibliography}
\end{document}